\documentclass[a4paper,11pt]{article}

\usepackage{jinstpub} 

\graphicspath{{figures/}}
\usepackage{multirow}
\usepackage{enumitem}
\usepackage{booktabs}
\usepackage{ulem}

\title{\boldmath Design and Performance Simulation of the Electromagnetic Calorimeter at EicC}

\author[a,1]{Ye Tian,\note{Corresponding author.}}
\author[a]{Souvik Maity,}
\author[a]{Jingyu Li,}
\author[a,b]{Yuancai Wu,}
\author[a]{Shan Sha,}
\author[a,b,c]{Yutie Liang,}
\author[a,b,c]{Aiqiang Guo,}
\author[a,b,c]{Yuxiang Zhao,}
\author[a,b,c,1]{and Dexu Lin}

\affiliation[a]{Institute of Modern Physics, Chinese Academy of Sciences, Lanzhou, Gansu Province 730000, China}
\affiliation[b]{School of Nuclear Science and Technology, University of Chinese Academy of Sciences, Beijing 100049, China}
\affiliation[c]{Heavy Ion Science and Technology Key Laboratory, Institute of Modern Physics, Chinese Academy of Sciences, Lanzhou 730000, China,}

\emailAdd{tianye@impcas.ac.cn}
\emailAdd{dxlin@impcas.ac.cn}

\abstract{The Electromagnetic Calorimeter (ECAL) is a key detector component for precise measurements of electrons and photons in electron–ion collision experiments. At the Electron–Ion Collider in China (EicC), high-performance calorimetry is essential for exploring the internal structure of nucleons and studying the dynamics of quarks and gluons within quantum chromodynamics (QCD). This paper presents the optimized design and performance simulation of the EicC ECAL system. The ECAL consists of three specialized sections tailored to distinct detection environments: (1) an electron-Endcap employing high-resolution pure Cesium Iodide (pCsI) crystals, (2) a central barrel, and (3) an ion-Endcap, both adopting a cost-effective Shashlik-style sampling calorimeter with improved light yield. Each segment’s geometry and material composition have been systematically optimized through Geant4 simulations to achieve excellent energy and position resolutions as well as strong electron–pion discrimination. The simulated performance indicates that the ECAL can achieve energy resolutions of 1.76\%/$\sqrt{E\text{(GeV)}}\oplus1.53\%$ for pCsI crystals and 5\%/$\sqrt{E\text{(GeV)}}\oplus4.2\%$ for Shashlik modules, meeting the design goals of the EicC detector. }

\keywords{Electromagnetic Calorimeter, Sampling Calorimeter, Homogeneous Calorimeter, Monte Carlo Simulation, Reconstruction}

\arxivnumber{} 

\begin{document}
\maketitle
\flushbottom
  

\section{Introduction}
The Electron–Ion Collider in China (EicC)~\cite{bib:eiccWhitePaper}, proposed as a new facility at the High Intensity Heavy-ion Accelerator Facility (HIAF)~\cite{YANG2013263} in Huizhou, is designed to advance our understanding of quantum chromodynamics (QCD). Through high-energy electron–ion collisions, it will probe the internal structure of nucleons, with particular emphasis on the origins of nucleon spin and mass, the dynamics of quarks and gluons, and the fundamental properties of the strong interaction.

\begin{figure}[htbp]
\centering
	\includegraphics[width=0.45\textwidth]{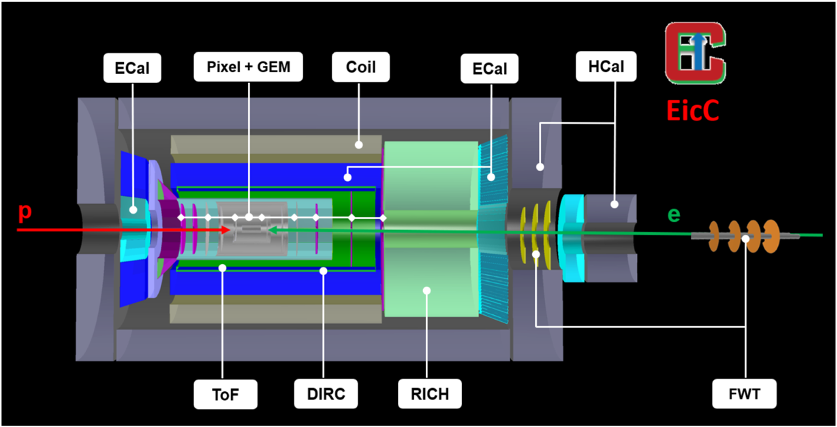}
	\includegraphics[width=0.45\textwidth]{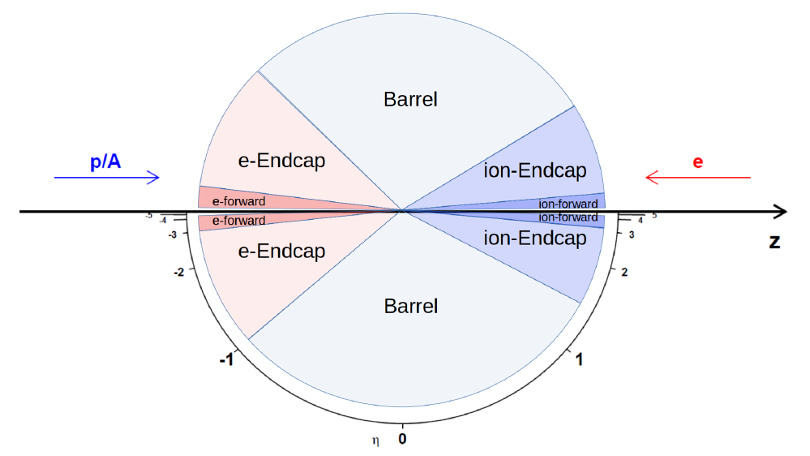}
	\caption{The EicC spectrometer design (left) and the coordinate definition of the detector segment (right)~\cite{bib:eiccWhitePaper}.}
	\label{eicc_detector}
\end{figure}

\begin{figure}[htbp]
\centering
	\includegraphics[width=0.45\textwidth]{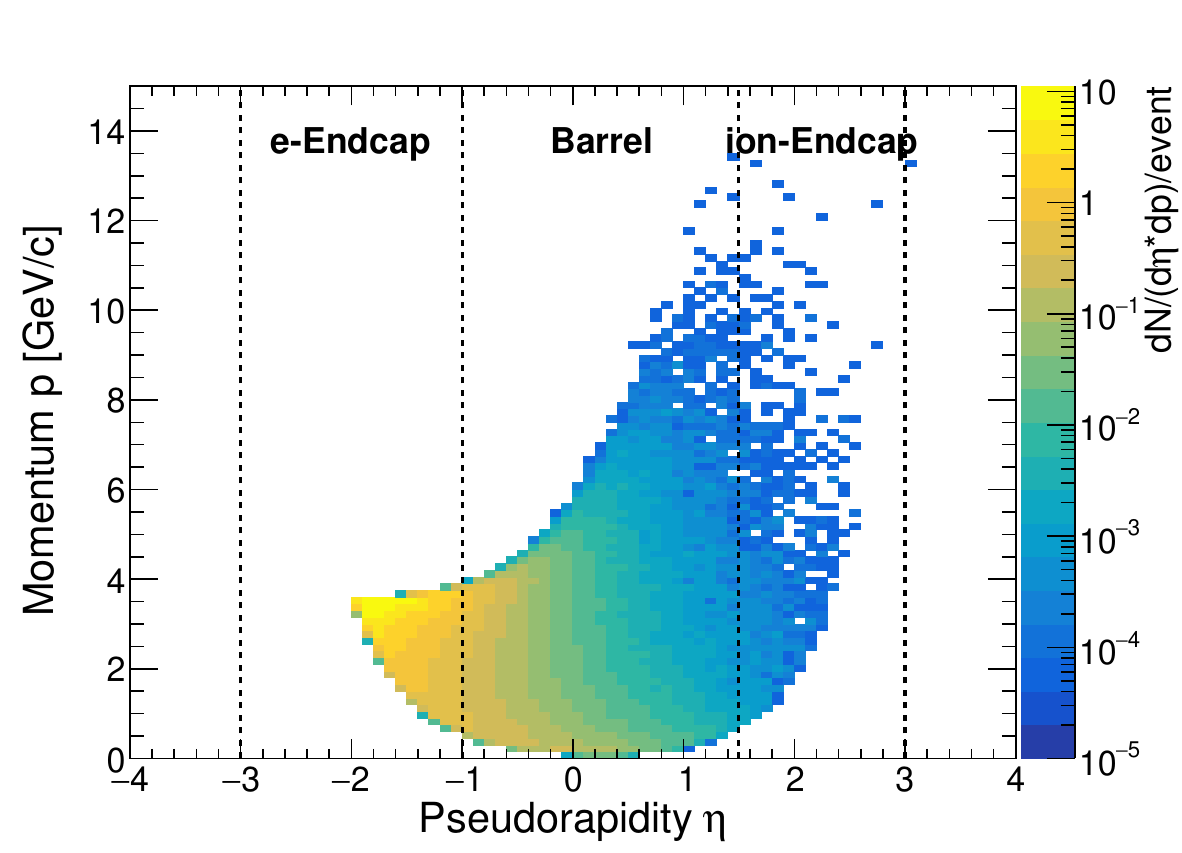}
	\includegraphics[width=0.45\textwidth]{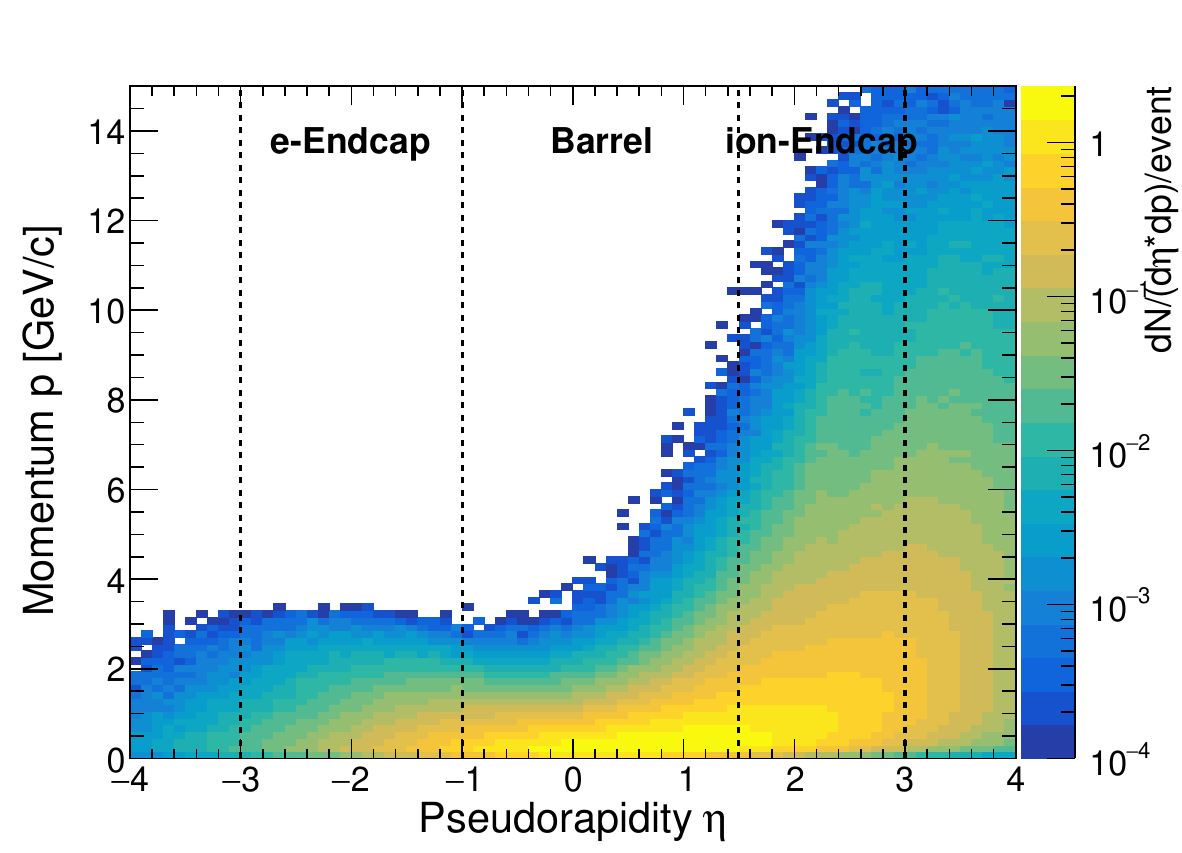}
    \caption{Momentum and pseudorapidity distributions of scattered electrons (left) and neutral pions (right) from EicC $ep$ scattering events. The data are obtained from Pythia simulations of 3.5~GeV electrons colliding with 20~GeV protons, with a kinematic requirement of $Q^{2} > 1~\text{GeV}^{2}$ applied.}
	\label{events_distribution}
\end{figure}

\begin{figure}[htbp]
\centering
	\includegraphics[width=0.45\textwidth]{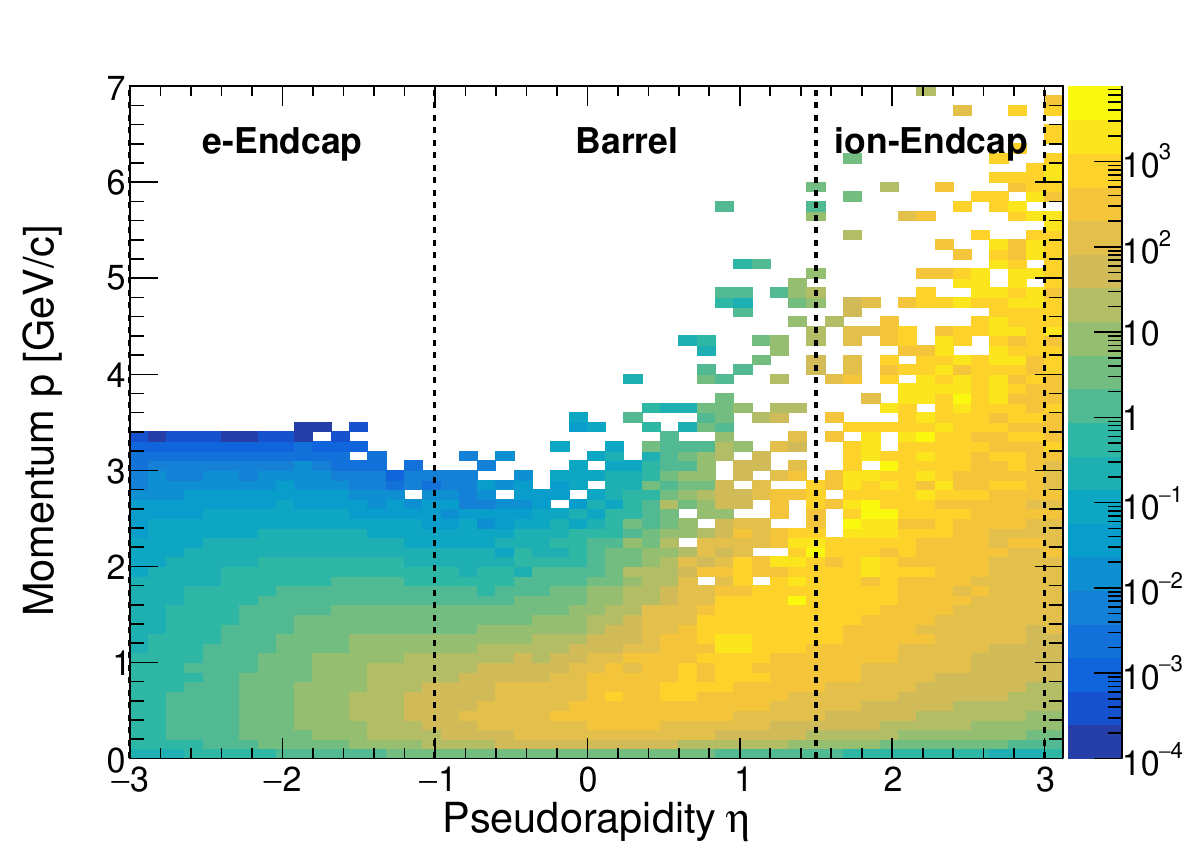}
    \caption{The $\pi/e$ ratio as a function of momentum and pseudorapidity is obtained from EicC $ep$ scattering events. The electrons are composed of scattered electrons and secondary electrons originating from $e^+e^-$ pair production. To evaluate the impact of pion background, in contrast to Fig.~\ref{events_distribution}, no kinematic requirement of $Q^{2} > 1~\mathrm{GeV}^{2}$ is applied in this case.}
    \label{pie_ratio}
\end{figure}

A schematic view of the EicC spectrometer is shown on the left of Fig.~\ref{eicc_detector}. The detector provides nearly $4\pi$ coverage for scattered particles. Fig.~\ref{events_distribution} shows the momentum and pseudorapidity distributions of scattered electrons and $\pi^{0}$ mesons, while Fig.~\ref{pie_ratio} presents the corresponding $\pi/e$ ratio. All figures are obtained from $ep$ scattering events generated with \textsc{Pythia}~\cite{pythia2020}, assuming 3.5~GeV electrons colliding with 20~GeV protons. The broad kinematic distributions highlight the need for uniform and high-resolution calorimetry across a wide angular range. The expected luminosity is $L = 4 \times 10^{33}\,\mathrm{cm^{-2}s^{-1}}$, corresponding to an interaction rate of approximately 83.2~kHz, with an average final-state particle multiplicity of about 9 per event.

Due to the asymmetric beam energies at the EicC, the ECAL must operate effectively across a wide and region-dependent dynamic range. The ECAL is responsible for detecting electrons and photons within the pseudorapidity range $|\eta| < 3$, which corresponds to approximately 99\% of the total solid angle. To achieve the required performance across this coverage, the ECAL is divided into three segments: the electron-endcap (e-Endcap), the central barrel, and the ion-Endcap. Each segment adopts a calorimeter technology optimized for its specific physics and geometrical constraints.

The main performance requirements of the ECAL are as follows:
\begin{itemize}[itemsep=0pt,parsep=0pt]
  \item Efficient reconstruction of scattered electrons and reliable separation from the pion background.
  \item Accurate reconstruction of photons with excellent energy and position resolutions.
  \item Capability to reconstruct high-energy $\pi^0$ mesons from their two-photon decays.
\end{itemize}

To meet these goals, the e-Endcap employs a high-resolution pure CsI~(pCsI) crystal calorimeter, targeting an energy resolution better than $2.5\%/\sqrt{E\text{(GeV)}}$ for electrons up to 4~GeV. The barrel and ion-Endcap regions utilize a Shashlik-style sampling calorimeter, optimized for light yield and uniformity, with an energy resolution of about $5\%/\sqrt{E\text{(GeV)}}$ for photons and electrons up to 15~GeV. A position resolution better than $6~\text{mm}/\sqrt{E\text{(GeV)}}$ and a timing precision below 1~ns are also required to suppress pile-up and improve particle identification.

Careful consideration has therefore been devoted to the design and optimization of each calorimeter segment. A combination of homogeneous crystal and sampling technologies provides a balanced solution between energy resolution, particle identification (PID), and cost, ensuring that the ECAL can fully support the precision physics program of the EicC.

\section{Electromagnetic Calorimeter Design}
Based on the detection requirements across different kinematic regions at EicC, the e-Endcap, barrel, and ion-Endcap sections of the ECAL are configured to cover pseudorapidity intervals of $(-3, -1)$, $(-1, 1.5)$, and $(1.5, 3)$, respectively. The segments and corresponding coverages are illustrated in Fig.~\ref{eicc_detector}. The e-Endcap is designed for precise detection of scattered electrons, utilizing a homogeneous crystal calorimeter with excellent energy resolution. For the calorimeter in barrel and ion-Endcap, where the energy resolution is not a critical factor and the e/$\pi$ identification performance will be a challenge, a Shashlik-style sampling calorimeter~\cite{bib:ecal:shashlik, Tian:2018lrh, Semenov_2020} is adopted as a compromise between the performance and the budget constraints.

\begin{figure}[htbp]
\centering
	\includegraphics[width=0.35\textwidth]{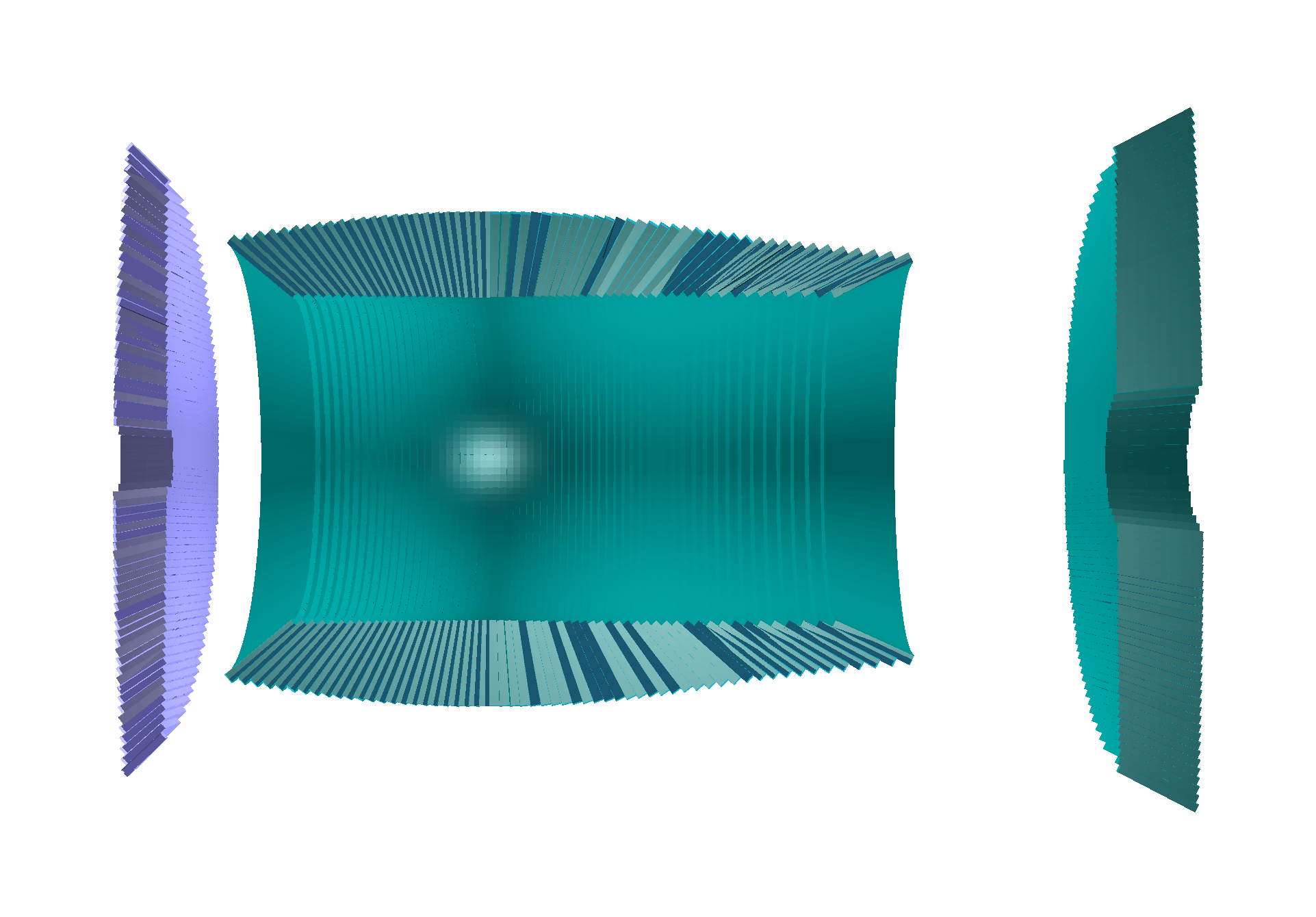}
	\includegraphics[width=0.45\textwidth]{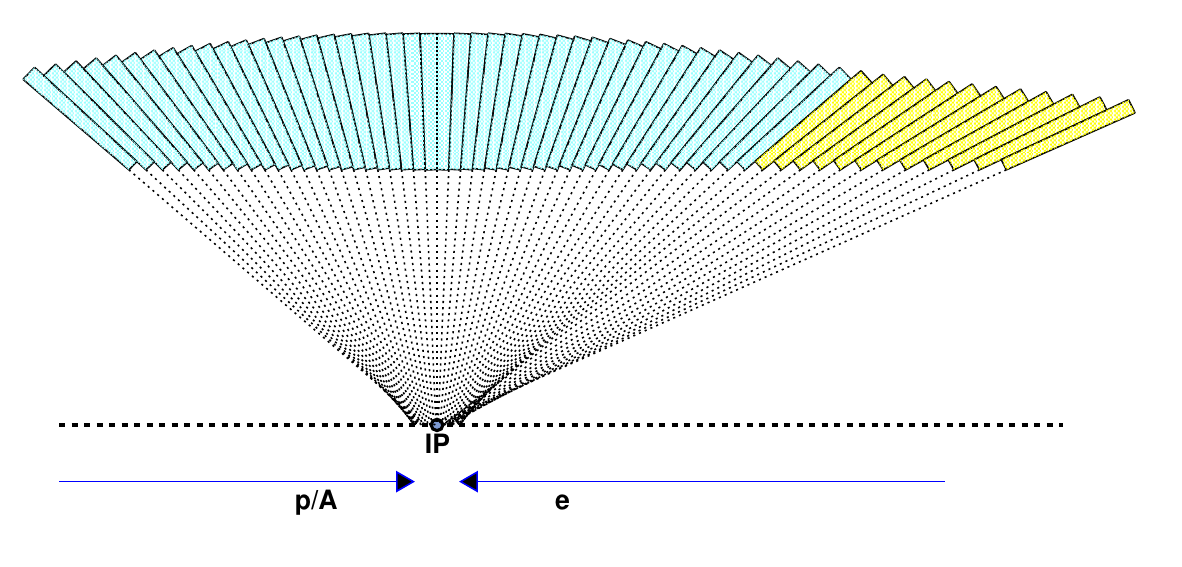}
	\caption{Overall design of the electromagnetic calorimeter (left), where the CsI crystal modules are shown in purple and the Shashlik modules in blue. Barrel module arrangement in a single row (right). Modules of the same color (blue or yellow) have identical geometry.}
	\label{emc_detector}
\end{figure}

The geometry of the ECAL is illustrated in Fig.~\ref{emc_detector}. The inner radius of the ECAL barrel is designed to be 0.9 m. Due to the differing measurement requirements between the electron and ion endcaps, the angular coverage in the barrel region is asymmetric, resulting in different barrel lengths on the electron and ion sides. The front planes of the e-Endcap and ion-Endcap are positioned at distances of 1.8~m and 3.5~m from the interaction point (IP), respectively. All ECAL modules are designed with a projective geometry and are oriented approximately toward the IP.

It should be noted that these geometric parameters are not yet finalized, as they remain highly dependent on the dimensions of the inner detector systems. A summary of the key parameters of the current ECAL design is provided in Table~\ref{table_ecal}.

\begin{table}
\centering
\begin{tabular}{lccc} 
\hline
Parameter                                     & \textbf{e-Endcap}                     & \textbf{Barrel} & \textbf{ion-Endcap}      \\ 
\hline
Material                                         & pCsI                                  & \multicolumn{2}{c}{Shashlik}               \\
Distance to IP [m]                            & 1.8                                   & 0.9 (radius)    & 3~                       \\
$\eta$ acceptance                             & [-3, -1]                              & [-1,  1.5]      & [1.5,  3]                \\
Inner radius [cm]                             & 18                                    & 90~             & 30                       \\
Outer radius [cm]                             & 153                                   & 150             & 145                      \\
Length [cm]                                   & 30 + 5 (readout)                       & \multicolumn{2}{c}{49 + 11 (readout)}       \\
Radiation length [$X_0$]               &  16  & \multicolumn{2}{c}{16}                                                            \\
Moli$\grave{e}$re radius [cm]                 & 3.5                                   & \multicolumn{2}{c}{5.02}                   \\
Front size [${\rm cm}^2$] & 4$\times$4           & \multicolumn{2}{c}{4$\times$4~}            \\
Rear size [${\rm cm}^2$] (max)  & 4.67$\times$4.67     & 5.8$\times$6.1~ & 4.6$\times$4.6~          \\
N layers                                      & -                                     & \multicolumn{2}{c}{240}                    \\
Scintillator thickness [mm]                   & -                                     & \multicolumn{2}{c}{1.5}                    \\
Lead thickness [mm]                           & -                                     & \multicolumn{2}{c}{0.35}                   \\
Reflector thickness [mm]                      & -                                     & \multicolumn{2}{c}{0.075}                  \\
Sampling ratio                                & -                                     & \multicolumn{2}{c}{0.33}                   \\
N fibers (front)                              & -                                     & \multicolumn{2}{c}{16}                     \\
Photon detector                               & APD or SiPM & \multicolumn{2}{c}{6$\times$6 ${\rm mm}^2$ SiPM}  \\
Total modules                                 & $\sim$3900              & $\sim$8000      & $\sim$3900               \\
\hline
\end{tabular}
\caption{The summary table of calorimeter design.}
\label{table_ecal}
\end{table}

In addition to the geometrical requirements, the expected energy coverage of the calorimeter is determined by the underlying event energy distributions, as previously shown in Fig.~\ref{events_distribution}. For the e-Endcap, the crystals are designed to measure energy deposits from 10 MeV up to 4 GeV. For the Shashlik calorimeters, the minimum detectable energy is about 25 MeV, while the maximum energy reaches about 10 GeV in the barrel region and up to 15 GeV in the ion-Endcap. These ranges define the required dynamic measurement capability of the detector and impose corresponding constraints on the detector design, including sensor response and signal readout.

\subsection{CsI Module Design}

For the R\&D of the e-Endcap, two types of homogeneous Cesium Iodide (CsI) crystal materials are under consideration: pure CsI (pCsI)~\cite{ATANOV2020162140} and thallium-doped CsI (CsI (Tl))~\cite{bes3}. Both materials offer comparable costs and exhibit excellent intrinsic energy resolution, typically around 2$\%/\sqrt{E({\rm GeV})}$. The main properties and differences between pCsI and CsI (Tl) are summarized in Table~\ref{csi_comp}.

\begin{table}[htbp]
\centering
\caption{The property comparison of pCsI and CsI (Tl)~\cite{pdg}. }
\begin{tabular}{lcc} 
\hline
\textbf{Parameters}             & \textbf{pCsI}                         & \textbf{CsI (Tl)}                       \\ 
\hline
Density [g/cm\textsuperscript{3}] & \multicolumn{2}{c}{4.51}                                     \\
Radiation length $X_0$ [cm]             & \multicolumn{2}{c}{1.86}                                     \\
Moli$\grave{e}$re radius [cm]               & \multicolumn{2}{c}{3.57}                                      \\
Refractive
  index (peak)         & 1.95                         & 1.79                          \\
Hygroscopicity                    & \multicolumn{1}{l}{Slightly} & \multicolumn{1}{l}{Slightly}  \\
Emission spectrum peak [nm]       & 310                          & 550                           \\
light yield rel. to NaI [\%]       & 5.6                          & 45                            \\
Decay time [ns]                   & 35                           & 1300                          \\
Absorption length of peak [cm]    & 100                           & 30                           \\
Preferred photon detector         & APD                          & PD                            \\
\hline
\end{tabular}
\label{csi_comp}
\end{table} 

The light yield of CsI (Tl) is approximately ten times higher than that of pCsI. However, simulation studies indicate that this increased light yield does not significantly improve the energy resolution, as the statistical uncertainty from photon counting is already sufficiently small. One drawback of CsI (Tl) is its long scintillation decay time, which could potentially lead to event pile-up, especially under high-rate conditions. Another limitation is its relatively short attenuation length, measured to be about 30~cm for photons with a wavelength of 550~nm~\cite{bib:ecal:csi_attenuation}, which can cause variations in photon collection efficiency depending on the depth of energy deposition within the crystal.

In contrast, pCsI provides superior timing performance owing to its much shorter scintillation decay time. However, its emission peak at 310~nm lies in the ultraviolet region, and its relatively low light yield necessitates the use of photodetectors with high quantum efficiency in the UV range. Taking these factors and the demand for precise timing capability into account, pCsI has been selected as the crystal material for the e-Endcap.

\begin{figure}[htbp]
\centering
	\includegraphics[width=0.4\textwidth]{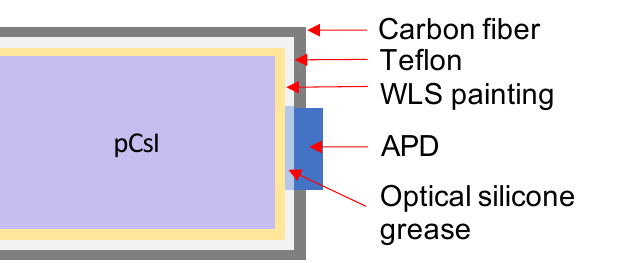}
	\caption{The pCsI module design.}
	\label{csi_module}
\end{figure}

The pCsI modules in the e-Endcap are designed with a projective trapezoidal geometry, featuring a total longitudinal length of 30~cm, equivalent to approximately 16$X_0$ radiation lengths. Each module has a lateral dimension of 4~$\times$~4~\text{cm}$^2$ at the front face, expanding to a maximum of about 4.7~$\times$~4.7~cm$^2$ at the rear face. Scintillation photons are collected by two large-area, UV-sensitive avalanche photodiodes (APDs), model Hamamatsu S8664-1010~\cite{bib:apd}, which offer a large dynamic range and relatively high quantum efficiency in the ultraviolet region. A schematic of the pCsI module design is shown in Fig.~\ref{csi_module}.

The light yield is enhanced by coating the surface of the pCsI crystal with the wavelength-shifting (WLS) material NOL9~\cite{bib:ecal:nol9}. NOL9 converts the ultraviolet scintillation light emitted by pCsI into yellow wavelengths, where the APDs exhibit higher quantum efficiency. The crystal module is further wrapped with Teflon tape, providing high reflectivity, particularly in the ultraviolet region. Optical silicone grease is applied between the APDs and the crystal surface to eliminate any air gap, as an air layer could cause total internal reflection and reduce photon collection efficiency. For mechanical stability and to maintain a light-tight enclosure, the entire assembly is housed within a carbon-fiber shell.

\subsection{Shashlik Module Design}
The Shashlik-style calorimeter is a sampling calorimeter comprising alternating layers of active and absorbing materials, with wavelength-shifting (WLS) fibers passing through each layer to collect scintillation light. It provides moderate energy resolution and has been adopted as a cost-effective solution for the ECAL calorimeters in both the barrel and ion-endcap regions. Recent beam tests have demonstrated that the Shashlik calorimeter can achieve an energy resolution of approximately 4.5\%/$\sqrt{E(\mathrm{GeV})}$~\cite{bib:ecal:nica_beam}, which is sufficient to meet the requirements for high-energy photon measurement and e/$\pi$ PID. The design parameters of the Shashlik calorimeter, including the sampling ratio, the number of layers, and the lateral cell size, can be adjusted to meet ECAL's specific performance requirements. For example, the longitudinal structure primarily affects the energy resolution, whereas the lateral dimension influences the position resolution.

\begin{figure}[htbp]
\centering
	\includegraphics[width=0.6\textwidth]{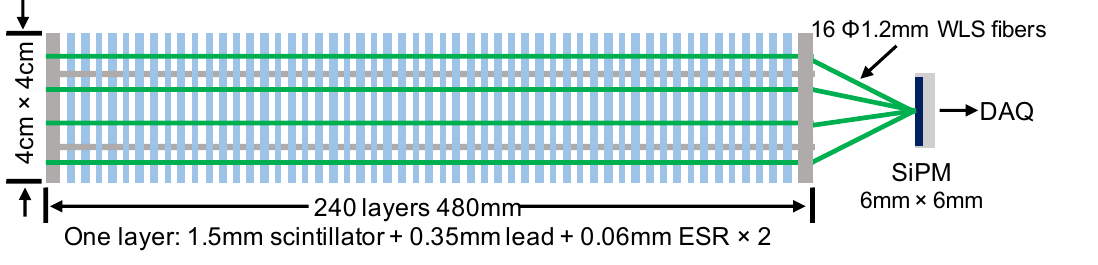}
	\includegraphics[width=0.2\textwidth]{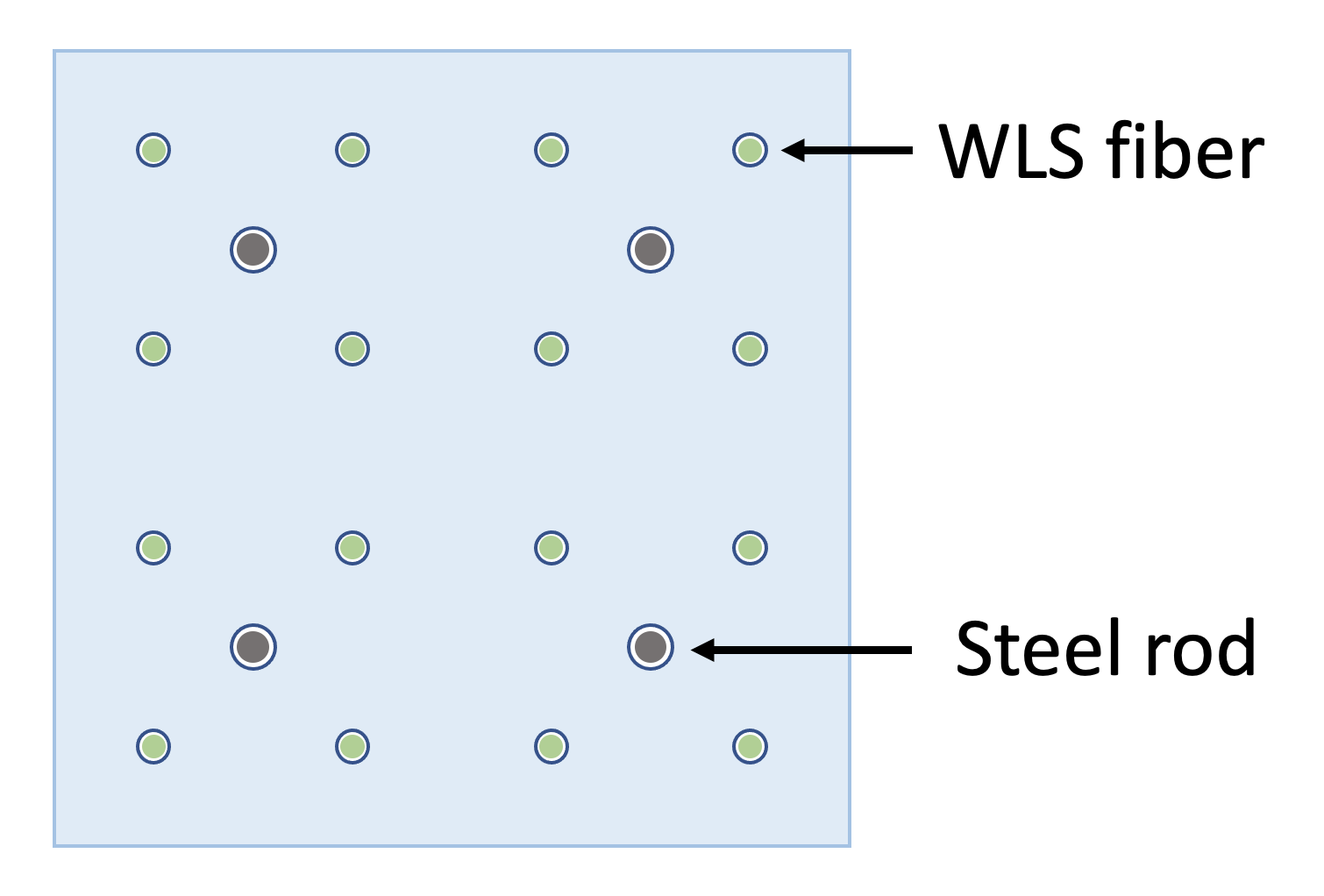}
	\caption{Longitudinal (left) and transverse (right) structural schematics of the Shashlik module.}
	\label{shashlik_module}
\end{figure}

In the current design, the alternating layers of the Shashlik module have been carefully optimized to balance performance requirements and engineering constraints. Based on simulation studies and experience from previous Shashlik calorimeter designs~\cite{bib:ecal:shashlik, Tian:2018lrh, Semenov_2020}, a configuration comprising 240 layers of 1.5~mm thick plastic scintillator as the active material and 0.35~mm thick lead plates as the absorber has been selected, the design of which has a sampling ratio of 0.33, as illustrated in Fig.~\ref{shashlik_module}. This configuration provides a total thickness of approximately 16 $X_0$, enabling effective photon detection for energies up to 15~GeV.

While simulation results indicate that increasing scintillator thickness and additional layers improve energy resolution through more complete energy deposit collection, these advantages are offset by longer module dimension and heightened light attenuation. The chosen configuration balances these opposing considerations to optimize detector performance and practical implementation.

The Shashlik module features a lateral size of 4~$\times$~4~cm$^2$. Each module is equipped with 16 double-cladding WLS fibers, with a diameter of 1.2~mm, which penetrate longitudinally through the holes in each layer to collect both scintillation and Cherenkov photons. The light collected by the WLS fibers is read out by a 6~$\times$~6~mm$^2$ SiPM mounted at one end of the module. To keep all layers firmly compressed together, four stainless-steel rods are used to apply adequate mechanical pressure.

To enhance the light yield, a 65~$\upmu$m thick 3M\texttrademark\ ESR mirror film, with a typical reflectivity greater than 98\% for visible light, is inserted between the scintillator and lead plates as a reflective layer. The ESR film is also attached to the distal end of each WLS fiber, serving as a mirror reflector. Additionally, the outer surface of the module is coated with TiO$_2$ paint, which enhances light collection and ensures optical isolation of the module. 

Simulation results indicate that a 1~GeV electron yields approximately 8000 photoelectrons at the SiPM readout. However, the actual light yield strongly depends on the light yield of the scintillator, the specific SiPM model employed, and the photon detection efficiency (PDE) at the chosen operating overvoltage. In practice, the overvoltage is often limited by the dynamic range of the readout electronics, which can lead to a reduced effective PDE.

The design of the Shashlik calorimeter in the barrel region faces a significant challenge: the rear end of each module is substantially larger than the front end due to the relatively long module length compared with the small barrel radius. The front face of each barrel module is positioned at a radius of 90~cm and has a lateral dimension of 4~$\times$~4~cm$^2$. Given a module length of 50~cm, the rear face expands to approximately 6.2~cm in width. This pronounced difference between the front and rear dimensions results in a non-uniform light collection efficiency along the module, which in turn can degrade the overall energy resolution. To mitigate this effect, additional WLS fibers are inserted at the rear end of each barrel module to enhance light collection uniformity. The resulting frustum-shaped design of the barrel Shashlik modules will be further studied and optimized in future work. 

\section{Simulation Results and Detector Design Optimization}
To evaluate the performance of the modules, Geant4~\cite{geant4} simulations were conducted using a 7~$\times$~7 module array for both the pCsI and Shashlik calorimeters. This configuration is sufficient to contain the majority of the lateral energy deposition from electromagnetic showers. A uniform square particle source with the same transverse size as a single module ($4 \times 4~\mathrm{cm}^2$) was used as input, allowing particles to hit the full module surface and thereby accounting for the overall impact of different incident positions on the detector performance. The simulations focus on both the energy deposition in sensitive materials and the photon transmission processes within the calorimeter. The evaluation and optimization of the detector design focus on several key performance metrics, including energy resolution, position resolution, and $e/\pi$ PID capability.

In the simulation of photon transmission processes, a detailed and realistic detector geometry is implemented in Geant4, including scintillation materials, reflective layers, wavelength-shifting materials (fibers and NOL9 painting), and optical sensors. Optical photon production and transport are fully simulated, covering scintillation light generation, reflection, absorption, wavelength shifting, and final photon detection. All material properties are taken from manufacturer datasheets or our experimental measurements, such as light yield, attenuation length, reflectivity, wavelength conversion efficiency, and photodetector quantum efficiency. Due to the large light yield and multiple internal reflections, the full optical simulation is computationally intensive and results in relatively low running efficiency.

A cluster reconstruction algorithm based on a neural-network and cellular-automata approach~\cite{bib:ecal_algorithm} is applied using the number of photoelectrons (NPE) information of individual modules. Adjacent modules with high NPE are grouped into a single cluster, while isolated modules with low energy deposits and no neighbours are discarded in the reconstruction procedure. This suppression of low-energy isolated hits reduces noise contributions but may introduce a small degradation in the energy resolution. This clustering algorithm is also employed in the $\pi^{0}$ reconstruction to identify two-photon candidates.

\subsection{Energy Reconstruction and Resolution}
The energy resolution of both the pCsI and Shashlik calorimeters at a fixed incident energy is obtained by fitting the energy deposition spectrum with a Gaussian function. However, for the pCsI crystal module, the energy deposition does not follow a purely Gaussian distribution due to the presence of asymmetric low-energy tails caused by energy leakage or photons escaping. Instead, the distribution is well described by the Crystal Ball function~\cite{crystal_ball}, which combines a Gaussian core with a power-law low-end tail. For the pCsI crystal, the energy resolution is approximated by the full width at half maximum (FWHM) divided by 2.35.

The energy resolution as a function of incident energy is fitted using a function with the electronic noise term ignored for the simulations~\cite{Fabjan2020}:
\begin{equation}
\frac{{\sigma}_E}{E} = \frac{a}{\sqrt{E({\rm GeV})}} \oplus c
\label{Eq_fittingE}
\end{equation}
where $a$ is the stochastic term, which dominates the energy resolution at low energies, and $c$ is the constant term, mainly arising from shower leakage (both lateral and longitudinal) within the finite $7\times7$ array, as well as non-uniform light collection among modules in the NPE-based results. The symbol $\oplus$ denotes a quadratic sum of the two components. 

\begin{figure}[htbp]
\centering
	\includegraphics[width=0.45\textwidth]{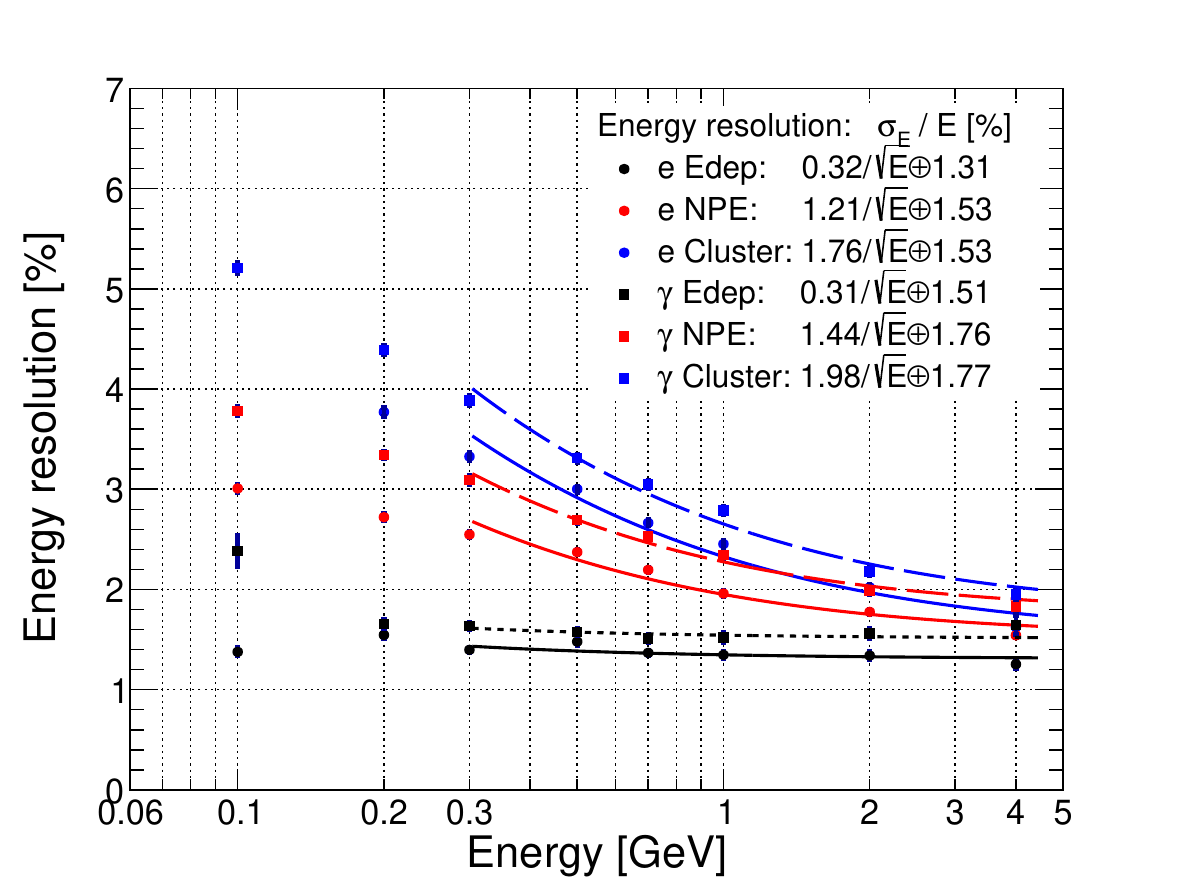}
	\includegraphics[width=0.45\textwidth]{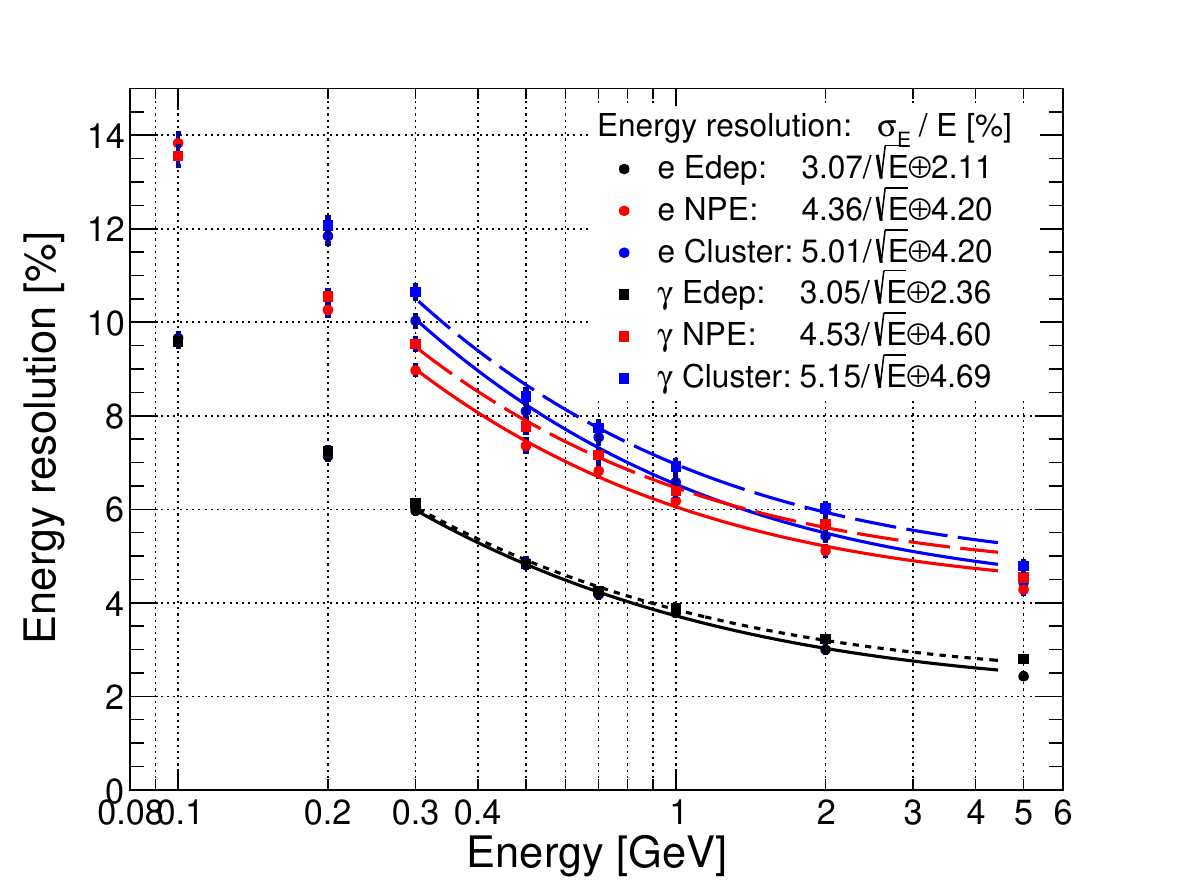}
	\caption{Simulation results of the energy resolution for electron and photon in the CsI (left) and Shashlik (right) module array.}
	\label{energy_reso}
\end{figure}

The simulated energy resolution results for both electrons and photons are shown in Fig.~\ref{energy_reso}, illustrating three stages of evaluation: the intrinsic energy deposition (Edep), the NPE statistics, and NPE statistics with applying the cluster reconstruction algorithm. The pCsI crystal calorimeter achieves an energy resolution of 1.76\%/$\sqrt{E}\oplus1.53\%$ for electrons, while the Shashlik calorimeter reaches 5\%/$\sqrt{E}\oplus4.2\%$. The photon energy resolution is slightly worse than that for electrons, primarily because photons have a smaller interaction cross section with matter, leading to a deeper shower start position and consequently a higher probability of energy leakage. For the Shashlik calorimeter, the energy resolution derived from NPE is significantly degraded compared to the Edep case, particularly in the constant term. This deterioration mainly arises from non-uniform light collection efficiency in the scintillator, especially its dependence on the transverse position of the energy deposition.

\subsection{Position Reconstruction and Resolution}
The hit position in the ECAL is reconstructed using a logarithmic energy-weighted centroid method~\cite{bib:ecal:position}, where the deposited energy (or NPE) in each module is used as the weight. Only modules above a certain energy threshold are included in the calculation. The reconstructed position is given by:

\begin{equation}
	X_{re} = \frac{\sum _{i}{w_i}{x_i}}{\sum _{i}{w_i}}, \  w_i = max\left\{ 0, W_0 +ln \left( \frac{E_i}{E_{total}} \right) \right\},
	\label{position_fitting}
\end{equation}

\noindent where $x_i$ ($y_i$) is the central coordinate of $i$-th module in the $x$ ($y$) direction, $w_i$ is the logarithmic weight factor, $E_i$ and $E_{total}$ is the energy deposit of single module and the sum of all modules, $W_0$ is a constant related to energy and shower shape and used as a threshold to discard the blocks with energy deposition below $E_{total}\cdot e^{-W_0}$.

\begin{figure}[htbp]
\centering
	\includegraphics[width=0.45\textwidth]{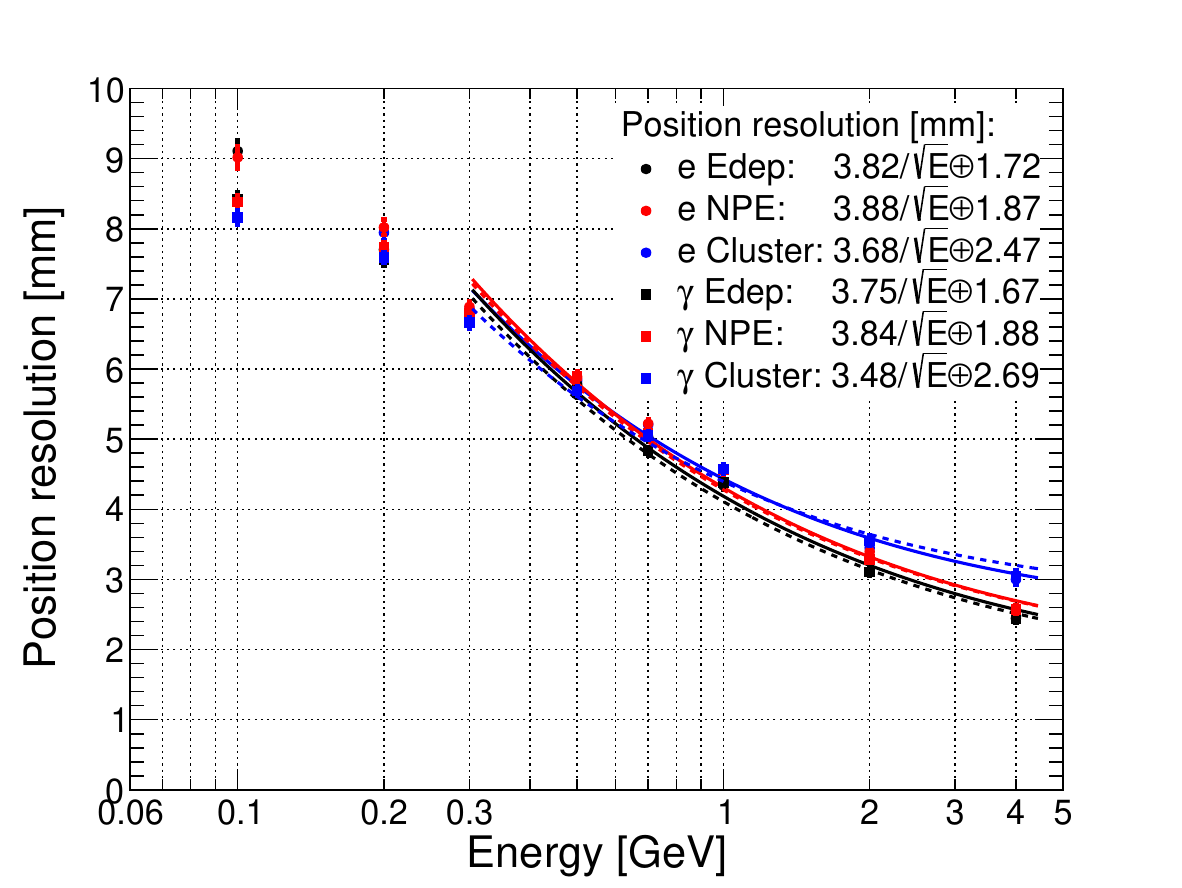}
	\includegraphics[width=0.45\textwidth]{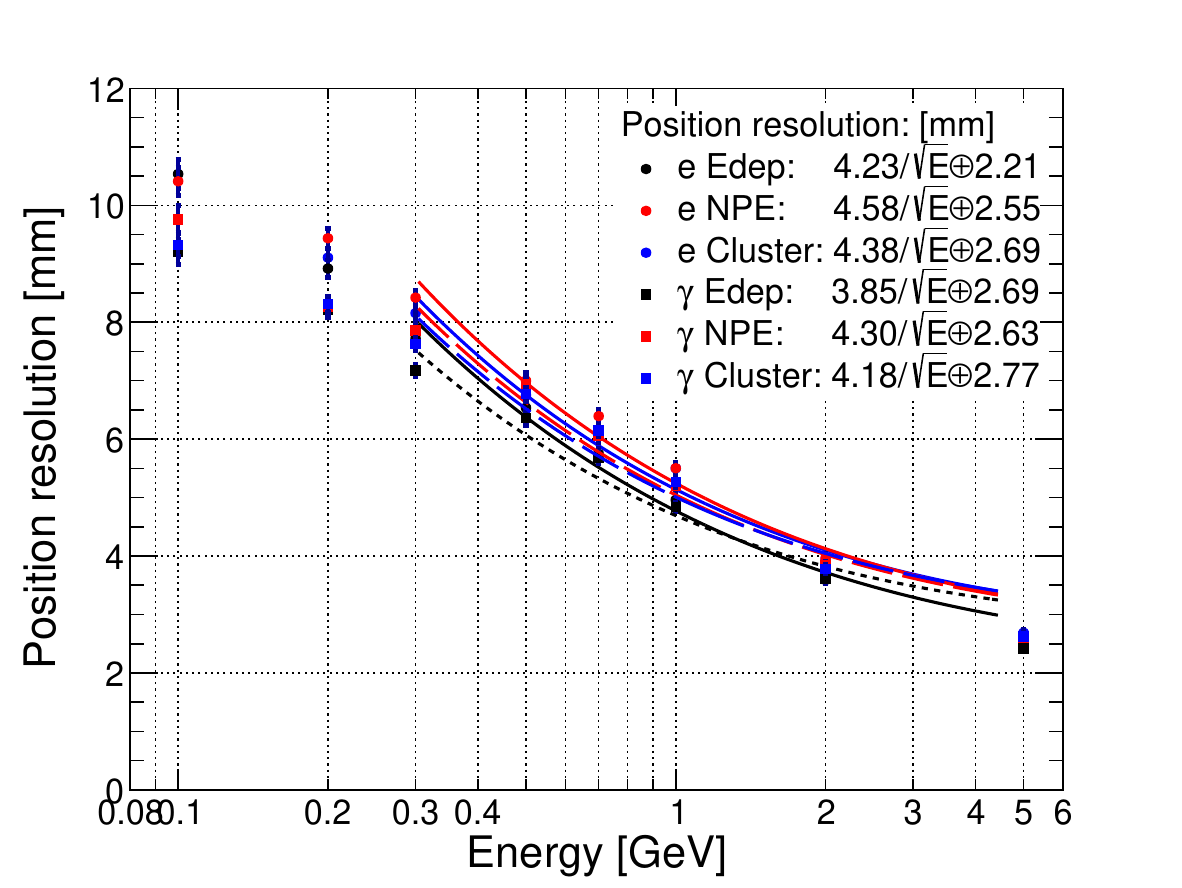}
 	\caption{Simulation results of the position resolution for electrons and photons in the CsI (left) and Shashlik (right) arrays, fitted with Eq.~(\ref{Eq_fittingE}) in the same way as the energy resolution.}
	\label{position_reso}
\end{figure}

Based on simulation studies of position reconstruction, the optimized values of $W_0$ are found to be 4.0 for the CsI calorimeter and 3.5 for the Shashlik calorimeter. The simulated position resolution results are shown in Fig.~\ref{position_reso}. The results indicate that the impact of energy resolution on position resolution is small. For both types of calorimeters with the same transverse size, the position resolution at 1 GeV is approximately 5~mm.

\subsection{Electron and Pion Identification}
One of the key performance requirements of the EicC ECAL is to provide additional hadron rejection capability, particularly for distinguishing pions from electrons, while simultaneously maintaining high efficiency for electron detection. Electrons typically deposit nearly all of their energy in the calorimeter via the development of an electromagnetic shower. In contrast, high-energy hadrons primarily behave as Minimum Ionizing Particles (MIPs), depositing only a fraction of their kinematic energy in the calorimeter. This difference can be exploited using the $E/p$ ratio, defined as the energy deposited in the calorimeter divided by the momentum measured by the tracking detector. For this simulation, the momentum $p$ is taken assuming ideal tracking resolution.

\begin{figure}[htbp]
    \centering
	\includegraphics[width=0.45\textwidth]{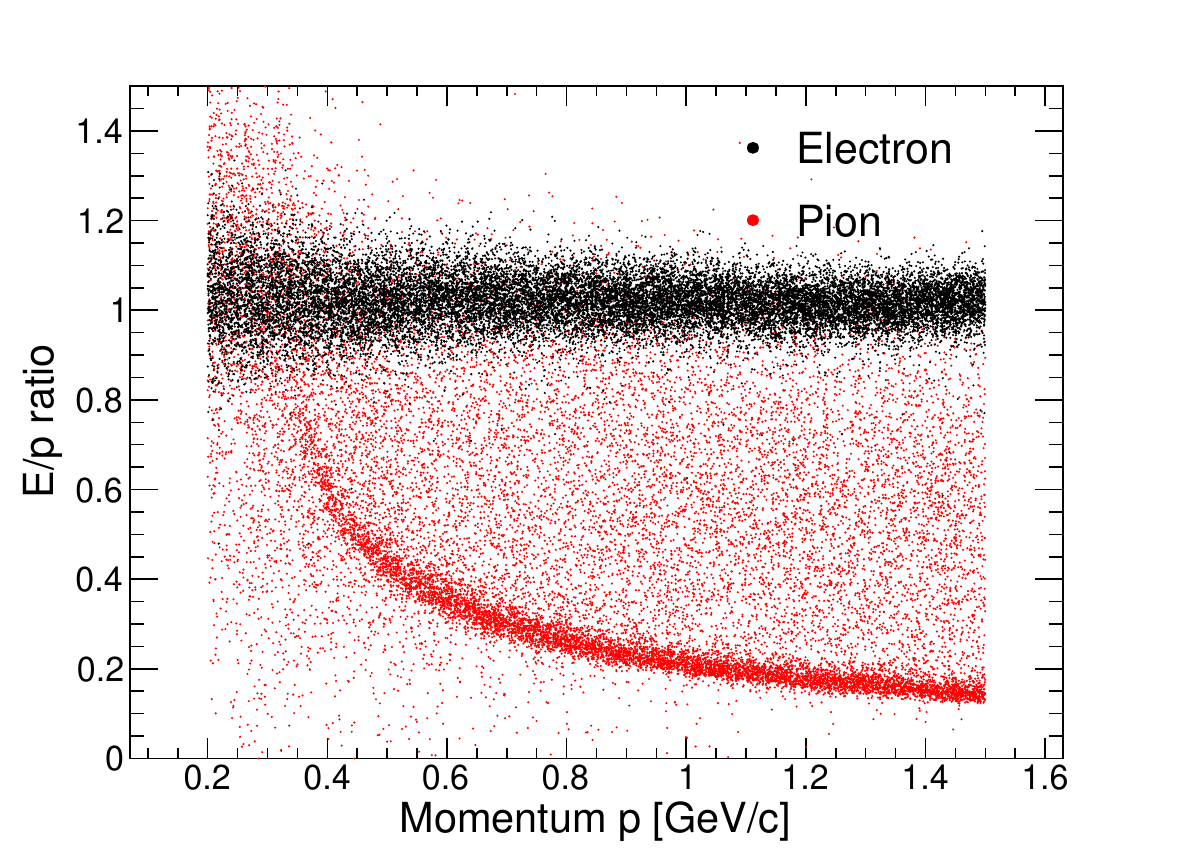}
	\includegraphics[width=0.45\textwidth]{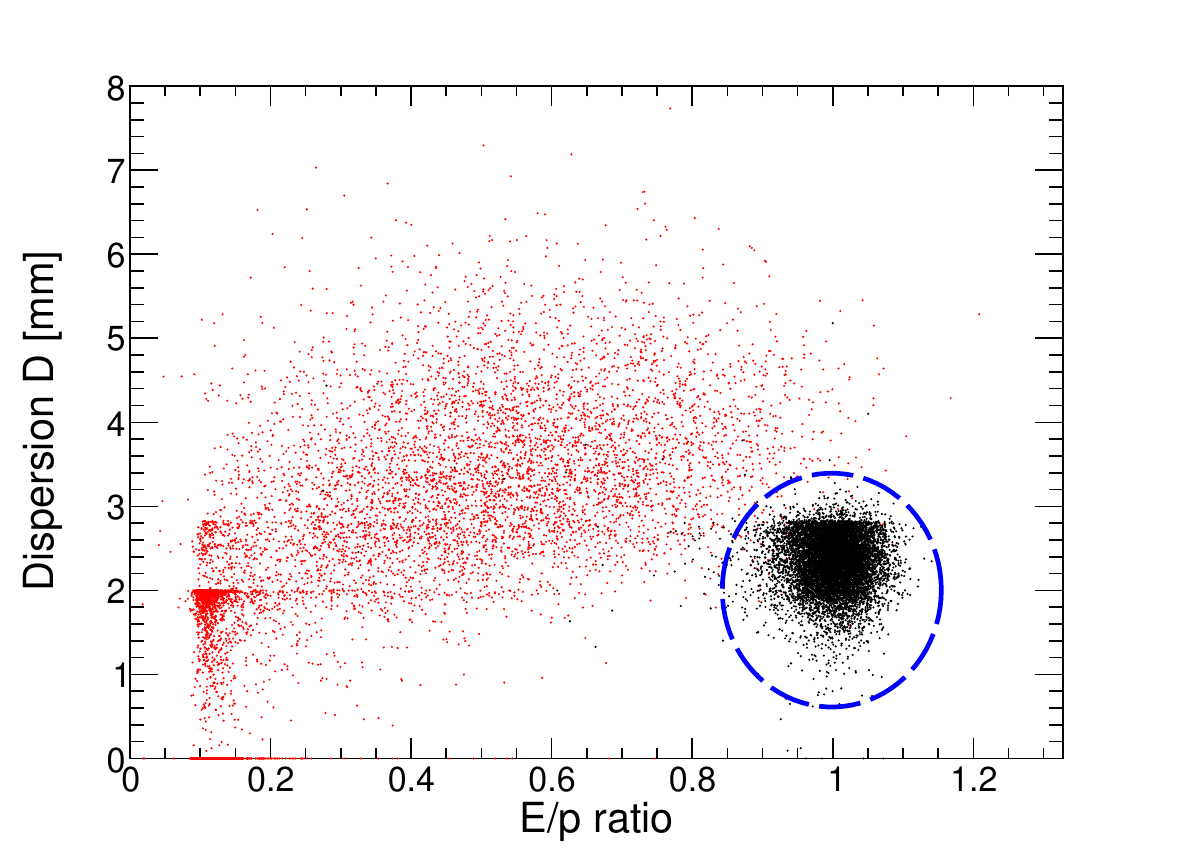}
    \caption{Simulation results of the $E/p$ ratio for electrons and pions at various momenta (left) and the electron–pion identification method based on the shower dispersion $D$ versus $E/p$ at 2~GeV/c(right). A circular requirement, shown in blue, is applied to distinguish electrons from the pion background.}
    \label{fig:pid}
\end{figure}

As shown in the left panel of Fig.~\ref{fig:pid}, the $E/p$ ratio for electrons is around the value 1, while most pions have a ratio less than 1. By applying an $E/p$ requirement, the majority of pion contamination can be suppressed, thereby improving electron purity. However, it should be noted that a fraction of pions can undergo hadronic interactions within the calorimeter and deposit energy up to their energy. In such cases, these pions mimic the response of electrons and can contaminate the electron events, particularly when the $\pi/e$ production ratio is large.

Additional discrimination between electrons and pions can be achieved by exploiting the difference in the spatial development of their showers. In general, hadronic showers tend to be broader and exhibit a larger lateral spread compared to the more compact electromagnetic showers induced by electrons. This characteristic can be quantified using the second moment of the hit position distribution, referred to as the shower dispersion $D$~\cite{bib:ecal:position}. The total dispersion $D$, which accounts for both the $x$ and $y$ directions, is defined as:

\begin{equation}
	D=\sqrt{D_x^2 + D_y^2}=\sqrt{\frac{\sum_i w_i (x_i-x_{\mathrm{re}})^2}{\sum_i w_i}+\frac{\sum_i w_i (y_i-y_{\mathrm{re}})^2}{\sum_i w_i}} 
	\label{shower_dispersion}
\end{equation}

\noindent Similar to Eq.~(\ref{position_fitting}) for position reconstruction, $x_i$ and $y_i$ denote the central coordinates of each module, $w_i$ represents the same logarithmic weight factor, and the $x_{\mathrm{re}}$ and $y_{\mathrm{re}}$ correspond to the reconstructed positions of the cluster.

Both the $E/p$ and dispersion ($D$) requirements are applied to achieve high electron efficiency while ensuring effective pion rejection. The right panel of Fig.~\ref{fig:pid} shows the distribution of dispersion $D$ versus the $E/p$ ratio for electrons and pions in the Shashlik array. The observed structures in $D$ reflect the distinct shower distributions: $D = 0$ corresponds to localized MIP-like pion events in a single module; $D \approx 2$ corresponds to energy sharing between two adjacent modules along a single transverse axis; and $D \approx 2.83$ ($\approx \sqrt{8}$) indicates the four-cell sharing pattern. A circular requirement is employed to optimize hadron rejection, effectively suppressing pion contamination while maintaining high electron detection efficiency. After applying the combined $E/p$ and $D$ requirements, the performance reaches a pion rejection factor of about 100:1 with an electron efficiency exceeding 99\%, as summarized in Table~\ref{PID_result}.

\begin{table}[htbp]
\centering
\caption{The PID performance for 2~GeV electrons and pions after applying the selection criteria shown in the right panel of Fig.~\ref{fig:pid}. Equal numbers of electrons and pions are simulated. Each column represents the PID efficiency: for example, in the first column, 99.25\% of electrons are correctly identified as electrons, while 0.75\% are misidentified as pions. The contamination rate is shown in the table rows, representing the fraction of misidentified particles in the final sample. For instance, in the electron sample, about $0.7/(99.25+0.7)\approx0.7\%$ of pions remain as contamination after the PID selection.}  
\begin{tabular}{c|c|c|c}
	\multicolumn{2}{c|}{\multirow{2}{*}{Percentage[\%]}} & \multicolumn{2}{c}{Real PID}  \\ 
	\cline{3-4}
	\multicolumn{2}{l|}{}                                & e     & $\pi$                 \\ 
	\hline
	\multirow{2}{*}{PID result} & e                      & 99.25 & 0.7                  \\ 
	\cline{2-4}
                            & $\pi$ & 0.75  & 99.3           
\end{tabular}
\label{PID_result}
\end{table}

\begin{figure}[htbp]
    \centering
	\includegraphics[width=0.45\textwidth]{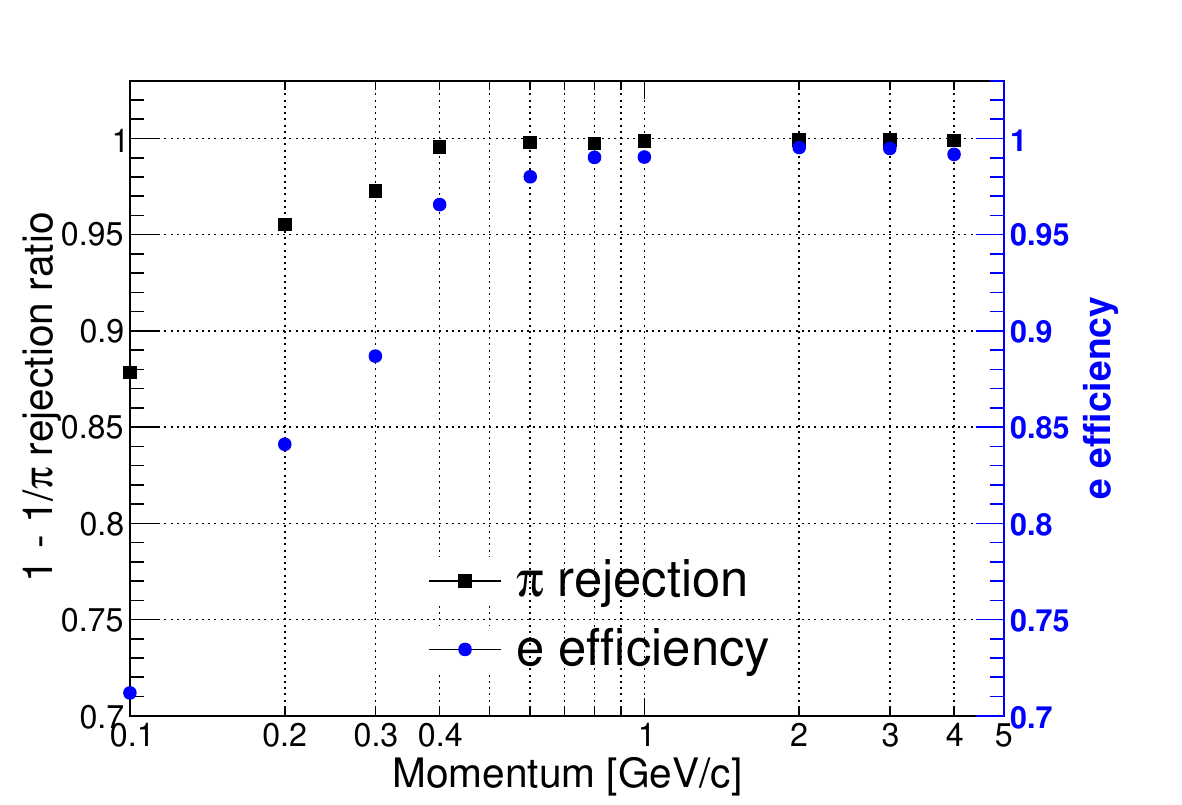}
	\includegraphics[width=0.45\textwidth]{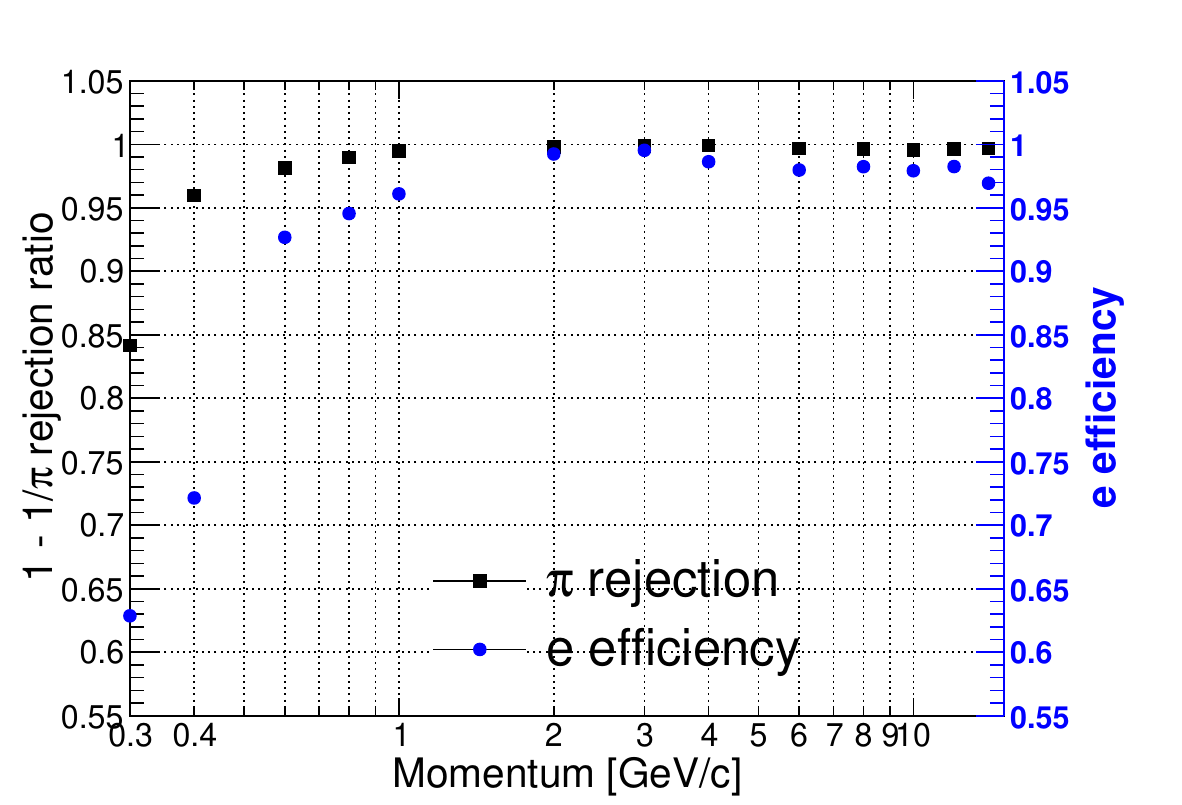}
    \caption{The pion rejection ratio and electron efficiency for different momenta of pCsI (left) and Shashlik (right) array.}
    \label{fig:pid_P}
\end{figure}

The Fig.\ref{fig:pid_P} shows the pion rejection and electron efficiency of the pCsI and Shashlik arrays at different momenta. To better approximate realistic experimental conditions, the $\pi/e$ ratio in the simulation is set to 10:1. It can be observed that both the electron efficiency and pion rejection generally improve with increasing energy. For the Shashlik calorimeter, however, the electron efficiency decreases slightly at higher energies. This is because high-energy pions increasingly exhibit electromagnetic shower-like behavior, making them more difficult to distinguish from electrons and degrading particle identification performance. 

In realistic experimental conditions, as shown in Fig.~\ref{pie_ratio}, the $\pi/e$ ratio can reach magnitudes as high as $5 \times 10^{3}$, which further highlights the stringent requirements on PID performance. Several approaches can be pursued to improve the PID capability of the calorimeter system. First, further optimization of the calorimeter-based PID algorithms is necessary. Second, improvements at the detector design level, including the introduction of a preshower detector~\cite{preshower} and a longitudinally segmented readout scheme~\cite{segmented_shashlik}, can provide more detailed information on the early stage and longitudinal development of particle showers. Finally, effective particle identification also relies on the combined use of other PID detectors, such as Cherenkov and Time-of-Flight (TOF) systems. Such a multi-detector approach is particularly important for enhancing e/$\pi$ discrimination in the low-momentum region.

\subsection{Design Optimization}
The calorimeter design was optimized through detailed simulations, with particular emphasis on energy resolution, which is strongly influenced by the longitudinal radiation length and sampling structure. For the pCsI module, performance can be tuned by adjusting the total crystal length, while the Shashlik module allows further flexibility by varying both the number and thickness of its alternating absorber–scintillator layers. In addition, the lateral size of the module plays a crucial role in determining position resolution, whereas the impact of energy resolution on position resolution is relatively minor. In this section, the Edep-based method was employed for both the simulation and the reconstruction in order to study the intrinsic performance of different design configurations.

\begin{figure}[htbp]
\centering
	\includegraphics[width=0.45\textwidth]{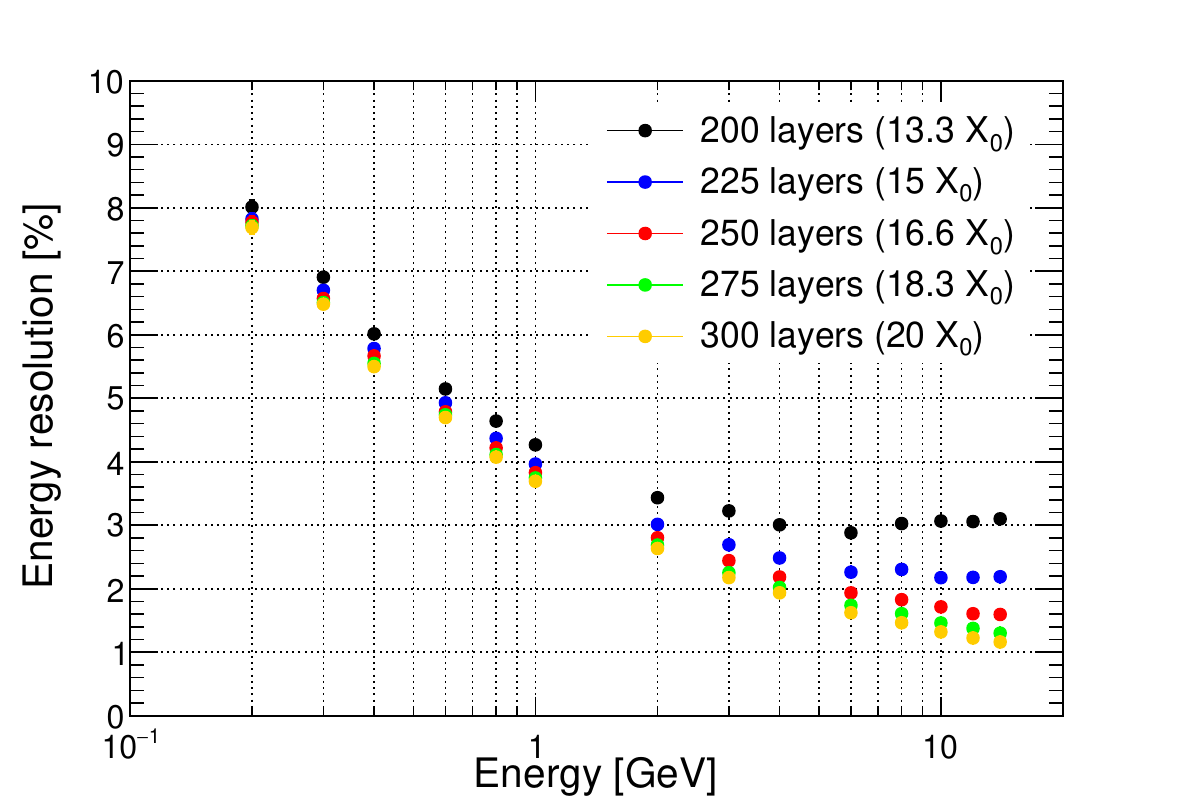}
	\includegraphics[width=0.45\textwidth]{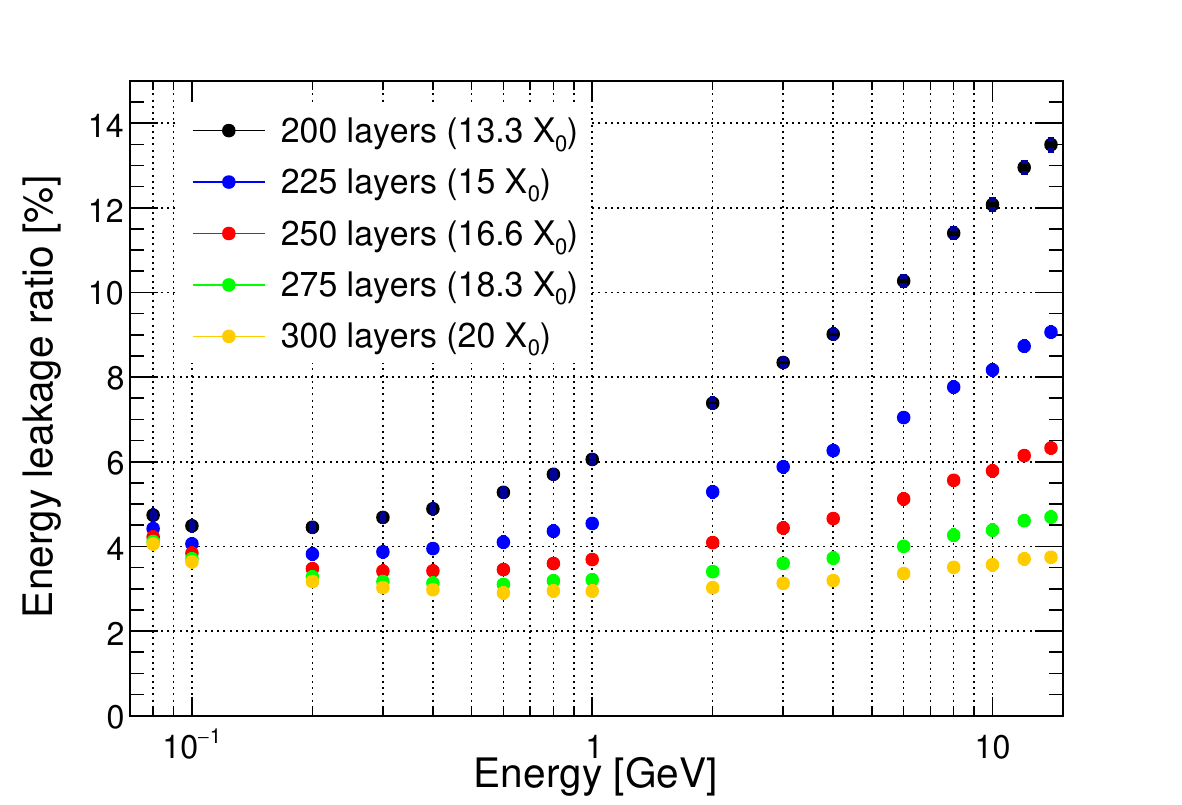}
	\caption{The energy resolution as a function of energy for different N layers (left). The energy resolution will get better with increasing layers. The energy leakage of different layers as a function of electron energy (right). }
	\label{horizontal_opti}
\end{figure}

As shown in Fig.~\ref{horizontal_opti}, increasing the total radiation length improves the energy resolution, especially for high-energy electrons. However, the module length is constrained by the available space in the barrel region and by cost considerations. After evaluating different configurations, a Shashlik module with 240 layers, corresponding to about 15.8~$X_0$, was chosen as a balanced solution between energy resolution and compactness. This configuration ensures adequate energy containment while maintaining manageable detector dimensions.

The thickness of the lead absorber layers also affects performance. Thicker absorber layers improve the energy resolution for high-energy electrons, which may benefit the ion-Endcap region where higher energies are expected. However, this comes at the cost of degraded resolution for low-energy photons, which are more sensitive to finer sampling. Therefore, the absorber thickness must be optimized according to the physics goals and energy spectrum of each detector segment, rather than adopting a single uniform design.

\begin{figure}[htbp]
\centering
	\includegraphics[width=0.45\textwidth]{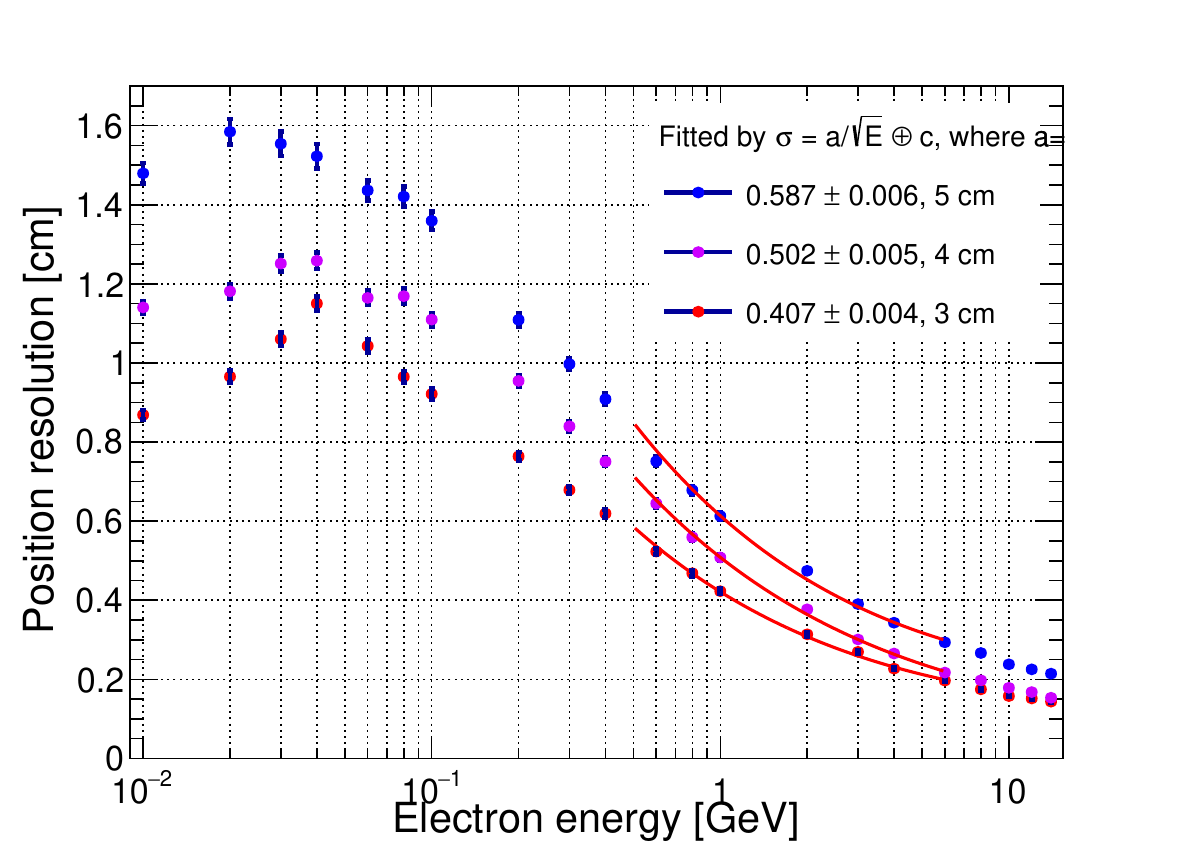}
	\caption{Position resolution as a function of energy for modules with lateral sizes of 3, 4, and 5~cm.}
	\label{lateral_opti}
\end{figure}

The lateral size of the module is another critical parameter. As indicated in Fig.~\ref{lateral_opti}, smaller modules yield better position resolution by offering finer spatial granularity. Nevertheless, reducing the module size increases the number of channels, which complicates the readout system and may introduce additional noise, affecting the shower reconstruction. A trade-off is therefore required between spatial resolution and system complexity.

\section{Neutral Pion Reconstruction}
The reconstruction of neutral pions ($\pi^{0}$) is an important physics objective and a key performance benchmark of the ECAL. As a primary product of hadronization in deep inelastic scattering, $\pi^{0}$ measurements provide valuable information on parton fragmentation and nucleon structure, particularly in semi-inclusive and exclusive processes. In addition, the well-known $\pi^{0}$ invariant mass serves as a standard reference for ECAL calibration and energy resolution validation. Therefore, efficient and precise $\pi^{0}$ reconstruction is both a physics requirement and a crucial indicator of the overall ECAL performance.

Since $\pi^0$ decays into two photons in about 99\% of cases, detecting both photons and separating those with small opening angles is crucial. The precision and efficiency of $\pi^0$ reconstruction depend on multiple factors, including energy and angular resolutions, detector acceptance, and shower separation capability. These factors reflect the combined performance of both the calorimeter and the reconstruction algorithm.

The distribution of $\pi^0$ mesons from EicC $ep$ collisions was shown in the right panel of Fig.~\ref{events_distribution}. Low-momentum $\pi^0$ events are abundant in the barrel region, while high-momentum events are concentrated in the ion-Endcap. These two regimes place different demands on detector design and reconstruction strategies.

In the $\pi^0$ rest frame, the two decay photons are emitted back-to-back with an opening angle of $180^\circ$. When the photon emission direction is perpendicular to the $\pi^0$ boost direction, the opening angle observed in the laboratory frame reaches its minimum. Assuming that the $\pi^0$ moves along the $z$ axis, the opening angle between the two photons can be expressed as
\begin{equation}
\theta_{\gamma\gamma} = 2\theta = 2\tan^{-1}\left(\frac{\sin\theta^{*}}{\gamma(\cos\theta^{*}+\beta)}\right)
\label{pi0_angle1}
\end{equation}

\noindent where $\theta$ is the angle between each photon and the $\pi^0$ momentum direction in the laboratory frame after Lorentz transformation, and $\theta^{*}$ is the photon emission angle in the rest frame. When $\theta^{*} = \pi/2$, both photons are perpendicular to the $\pi^0$ flight direction, and in this case the opening angle reaches its minimum value:

\begin{equation}
\theta_{\gamma\gamma}^{min} = 2\tan^{-1}\left(\frac{1}{\gamma\beta}\right) =  2\tan^{-1}\left(\frac{m_{\pi^{0}}}{p_{\pi^{0}}}\right).
\label{pi0_angle2}
\end{equation}

\begin{figure}[htbp]
\centering
	\includegraphics[width=0.45\textwidth]{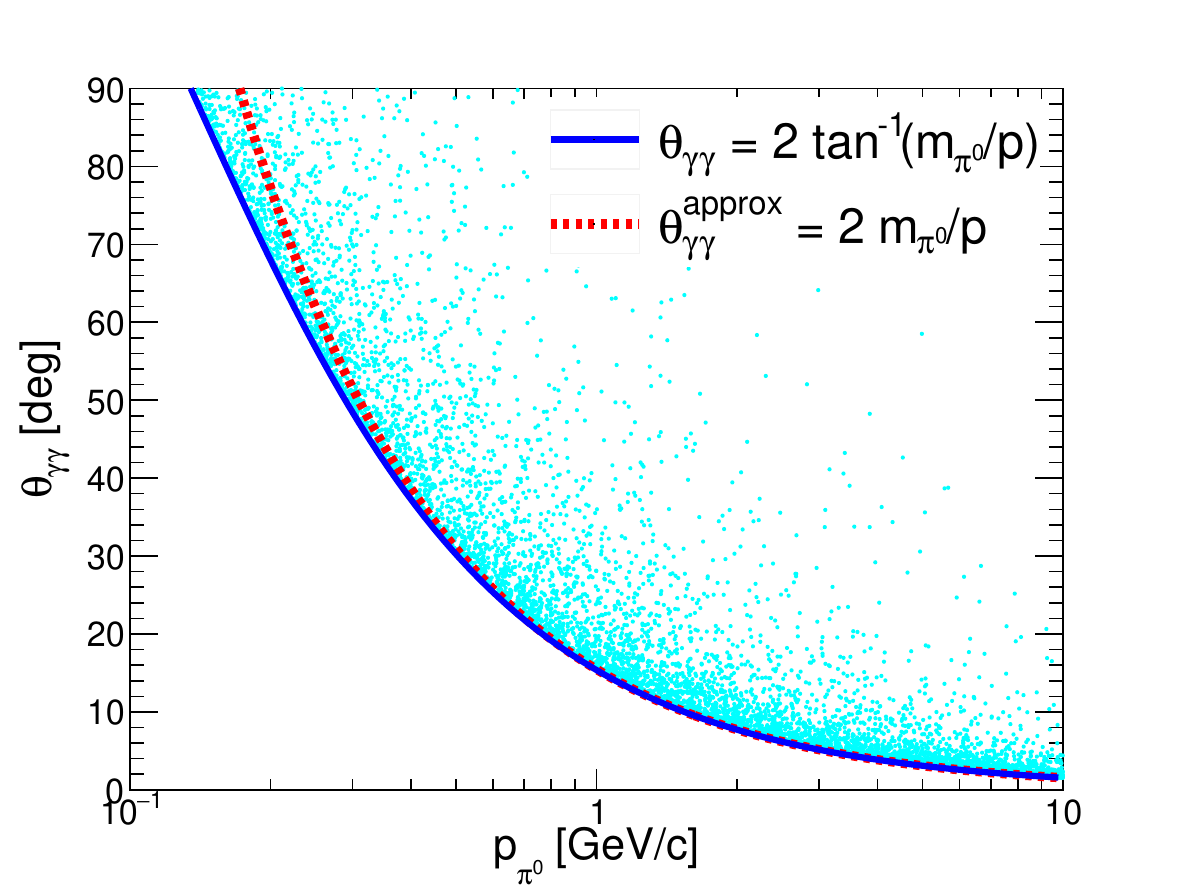}
	\caption{Angular distributions between the two decay photons from $\pi^0$ mesons with different momenta obtained from Geant4 simulation.}
	\label{pi0_minimum_angle}
\end{figure}

Fig.~\ref{pi0_minimum_angle} shows the distribution of the opening angle between the two decay photons for different $\pi^0$ momenta, obtained from Geant4 simulations, which is found to be consistent with Eq.~(\ref{pi0_angle2}). In the relativistic limit, where the $\pi^0$ momentum is much greater than its rest mass ($p_{\pi^0} \gg m_{\pi^0}$), Eq.~(\ref{pi0_angle2}) can be approximated by $\theta_{\gamma\gamma} \approx 2m_{\pi^0}/p_{\pi^0}$. This approximation clearly illustrates that the opening angle between the two photons decreases rapidly with increasing energy of $\pi^0$, leading to a higher probability of cluster overlap in the calorimeter. This effect is particularly significant in the ion-Endcap region, where high-momentum $\pi^0$ mesons are predominant. To mitigate the overlap and improve the reconstruction efficiency, the calorimeter geometry can be optimized by increasing the distance between the ion-Endcap and the interaction point and by refining the cluster reconstruction algorithms.

\begin{figure}[htbp]
\centering
	\includegraphics[width=0.45\textwidth]{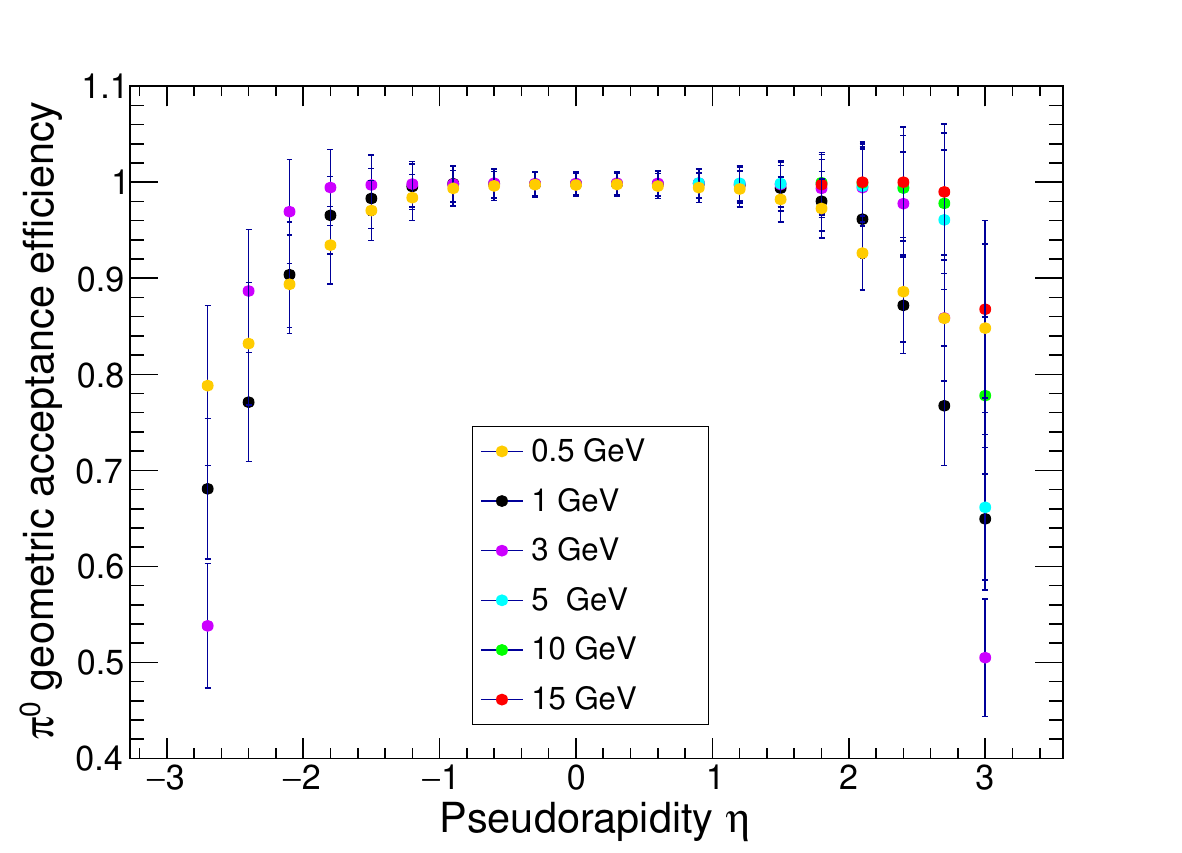}
	\includegraphics[width=0.45\textwidth]{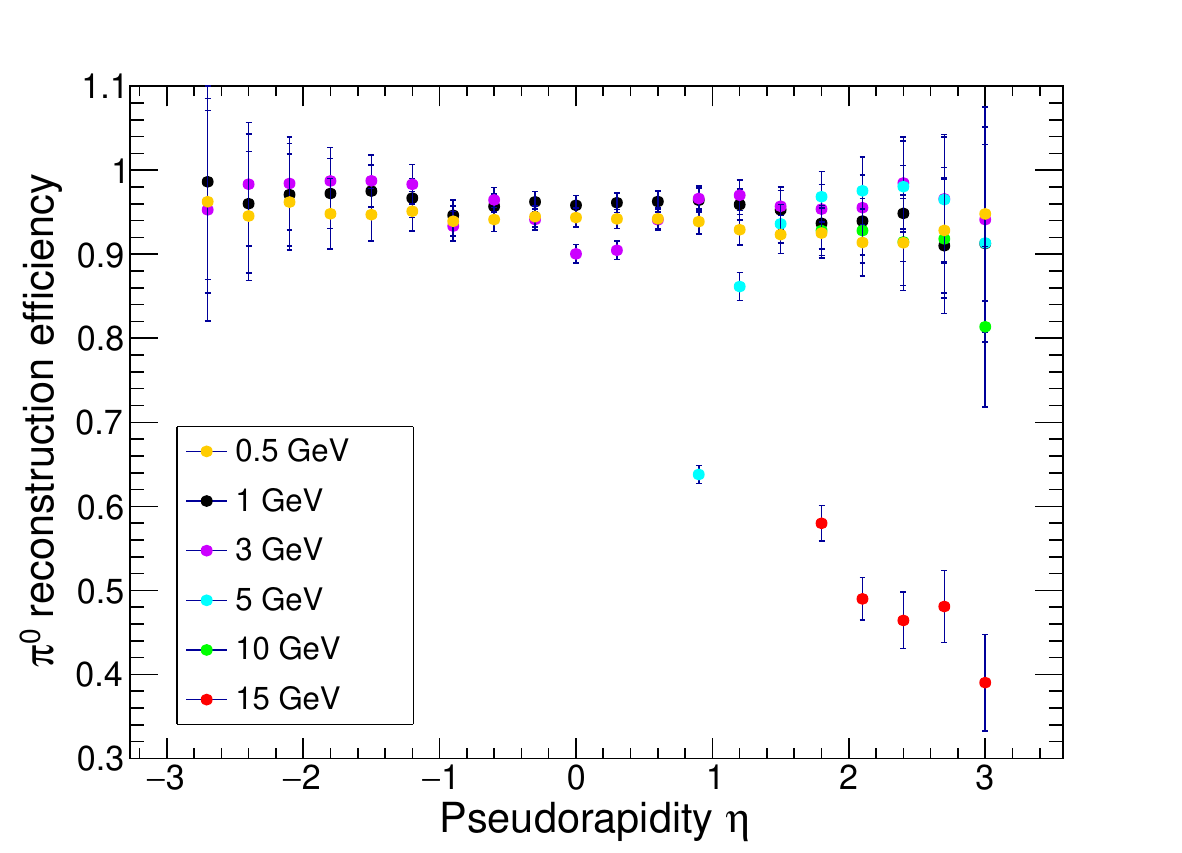}
	\caption{The $\pi^{0}$ geometric acceptance efficiency as a function of pseudorapidity for different momenta (left), and the $\pi^{0}$ reconstruction efficiency (right).}
	\label{pi0_effi}
\end{figure}

Simulation results using a more realistic geometry, as illustrated in the left panel of Fig.~\ref{emc_detector}, indicate three main factors that can degrade the $\pi^0$ reconstruction efficiency. First, the energy of each decay photon must exceed the electronic threshold to suppress noise and ensure reliable detection. In this study, each photon cluster is required to satisfy a minimum energy threshold of $10$~MeV. Second, one of the photons may escape through the beam hole or hit the detector edge, resulting in a loss of geometric acceptance. The geometric acceptance efficiency is defined as the fraction of generated $\pi^0$ events for which both decay photons fall within the ECAL acceptance, as illustrated in the left panel of Fig.~\ref{pi0_effi}.

Third, when the opening angle between the two photons is small, their electromagnetic showers may overlap, leading to the reconstruction of a single cluster. The reconstruction efficiency is defined as the fraction of events in which two photon clusters are reconstructed in the ECAL and their invariant mass falls within the $110$--$160$~MeV window, relative to the number of events satisfying the geometric acceptance requirement. As shown in the right panel of Fig.~\ref{pi0_effi}, high-momentum $\pi^0$ mesons exhibit lower reconstruction efficiency. Improving the angular resolution, for example by increasing the distance between the detector and the IP, together with optimizing the reconstruction algorithm, can effectively mitigate this inefficiency.

\section{Conclusions}
A comprehensive design and optimization study of the ECAL for the EicC has been presented, combining two complementary technologies to meet the distinct physics and geometric requirements of the detector: a pCsI crystal calorimeter in the e-Endcap and Shashlik-style sampling calorimeters in the barrel and ion-Endcap regions.

Detailed Geant4 simulations, incorporating realistic detector responses, show that the proposed configuration meets the performance targets required by the EicC physics program. The pCsI modules offer excellent energy resolution of $1.76\%/\sqrt{E\text{(GeV)}}\oplus1.53\%$, superior position resolution, and strong timing performance, benefiting from their enhanced effective light yield. The Shashlik modules, optimized in sampling structure, radiation length, and light collection uniformity, provide energy and position resolutions of $5\%/\sqrt{E\text{(GeV)}}\oplus4.2\%$ and 5~mm at 1~GeV, respectively, along with effective electron–pion separation. 

These results indicate that the current ECAL design can reliably support precision measurements of electrons and photons across the full acceptance of the EicC detector. Future work will focus on detailed digitization modeling, prototype construction and validation, and advancing the ECAL toward full engineering design and construction.

\section*{Acknowledgments}{
\sloppy
This work is supported by National Key R\&D Program of China under Contract
No.~2023YFA1606903, CAS Pioneer Hundred Talents Program, and Postdoc Program of Department of Human Resources and Social Security of Gansu Province.
\par}

\bibliographystyle{JHEP}
\bibliography{references}

@article{bib:eiccWhitePaper,
author = {Daniele P. Anderle and Valerio Bertone and Xu Cao and others},
title = {Electron-ion collider in {China}},
publisher = {Front. Phys. },
year = {2021},
journal = {Front. Phys.},
volume = {16},
number = {6},
eid = {64701},
numpages = {0},
pages = {64701},
url = {https://journal.hep.com.cn/fop/EN/abstract/article_29192.shtml},
doi = {10.1007/s11467-021-1062-0}
}

@article{YANG2013263,
title = {High {Intensity} heavy ion {Accelerator} {Facility} ({HIAF}) in {China}},
journal = {Nucl. Instr. Meth. Phys. Res. B},
volume = {317},
pages = {263-265},
year = {2013},
note = {XVIth International Conference on ElectroMagnetic Isotope Separators and Techniques Related to their Applications, December 2–7, 2012 at Matsue, Japan},
issn = {0168-583X},
doi = {10.1016/j.nimb.2013.08.046},
url = {https://www.sciencedirect.com/science/article/pii/S0168583X13009877},
author = {J.C. Yang and others},
}

@article{pythia2020,
title = {The {Pythia} event generator: Past, present and future},
journal = {Comput. Phys. Commun.},
volume = {246},
pages = {106910},
year = {2020},
issn = {0010-4655},
doi = {10.1016/j.cpc.2019.106910},
url = {https://www.sciencedirect.com/science/article/pii/S0010465519302899},
author = {Torbjörn Sjöstrand},
}

@article{bib:ecal:shashlik,
title = {An improved {Shashlyk} calorimeter},
journal = {Nucl. Instr. Meth. Phys. Res. A},
volume = {584},
number = {2},
pages = {291-303},
year = {2008},
issn = {0168-9002},
doi = {10.1016/j.nima.2007.10.022},
url = {https://www.sciencedirect.com/science/article/pii/S0168900207021717},
author = {G.S. Atoian and G.I. Britvich and S.K. Chernichenko and others},
}

@article{bib:ecal:csi_attenuation,
title = {Simulations of light collection in long tapered {CsI(Tl)} scintillators using real crystal surface data and comparisons to measurement},
journal = {Nucl. Instr. Meth. Phys. Res. A},
volume = {1003},
pages = {165302},
year = {2021},
issn = {0168-9002},
doi = {10.1016/j.nima.2021.165302},
url = {https://www.sciencedirect.com/science/article/pii/S0168900221002862},
author = {A. Knyazev and J. Park and P. Golubev and others},
}

@article{bib:ecal:nica_beam,
title = {Beam test results of two shashlyk {ECal} modules for {NICA-MPD}},
journal = {Nucl. Instr. Meth. Phys. Res. A},
volume = {958},
pages = {162833},
year = {2020},
note = {Proceedings of the Vienna Conference on Instrumentation 2019},
issn = {0168-9002},
doi = {10.1016/j.nima.2019.162833},
url = {https://www.sciencedirect.com/science/article/pii/S0168900219312689},
author = {Yulei Li and Dong Han and Yi Wang and others},
}

@article{bib:ecal:position,
title = "{A simple method of shower localization and identification in laterally segmented calorimeters}",
journal = {Nucl. Instr. Meth. Phys. Res. A},
volume = {311},
number = {1},
pages = {130-138},
year = {1992},
issn = {0168-9002},
doi = {10.1016/0168-9002(92)90858-2},
url = {https://www.sciencedirect.com/science/article/pii/0168900292908582},
author = {T.C. Awes and F.E. Obenshain and F. Plasil and others},
}

@article{Semenov_2020,
doi = {10.1088/1748-0221/15/05/C05017},
url = {https://doi.org/10.1088/1748-0221/15/05/C05017},
year = {2020},
month = {may},
publisher = {},
volume = {15},
number = {05},
pages = {C05017},
author = {Semenov, A.Yu. and others},
title = "{Electromagnetic calorimeter for MPD spectrometer at NICA collider}",
journal = {JINST},
}

@article{bib:ecal_algorithm,
title = "Application of neural networks and cellular automata to interpretation of calorimeter data",
journal = "Nucl. Instr. Meth. Phys. Res. A",
volume = "362",
number = "2",
pages = "478 - 486",
year = "1995",
issn = "0168-9002",
doi = "10.1016/0168-9002(95)00217-0",
url = "http://www.sciencedirect.com/science/article/pii/0168900295002170",
author = "V. Breton and H. Fonvieille and P. Grenier and C. Guicheney and J. Jousset and Y. Roblin and F. Tamin"
}

@article{bib:ecal:nol9,
title = {Highly {Efficiency} {Wavelength} {Shifters}: {Design}, {Properties}, and {Applications}},
journal = {INEOS OPEN},
volume = {2},
number = {4},
pages = {112-123},
year = {2019},
doi = {10.32931/io1916r},
url = {http://ineosopen.org/f/io1916r.pdf},
author = {O. V. Borshchev and N. M. Surin and M. S. Skorotetcky and S. A. Ponomarenko}
}

@article{Tian:2018lrh,
    author = "Tian, Y. and Chen, J. P. and Feng, C. and Jiao, J. and Li, A. and Yu, Y. and Zheng, X.",
    editor = "Liu, Zhen-An",
    title = "{Electromagnetic Calorimeter Prototype for the SoLID Project at Jefferson Lab}",
    doi = "10.1007/978-981-13-1316-5_15",
    journal = "Springer Proc. Phys.",
    volume = "213",
    pages = "80--85",
    year = "2018"
}

@article{ATANOV2020162140,
title = "{Design and status of the Mu2e crystal calorimeter}",
journal = {Nucl. Instr. Meth. Phys. Res. A},
volume = {958},
pages = {162140},
year = {2020},
note = {Proceedings of the Vienna Conference on Instrumentation 2019},
issn = {0168-9002},
doi = {10.1016/j.nima.2019.04.094},
url = {https://www.sciencedirect.com/science/article/pii/S0168900219305832},
author = {N. Atanov and V. Baranov and C. Bloise and J. Budagov and F. Cervelli and S. Ceravolo and F. Colao and M. Cordelli and G. Corradi and Yu.I. Davydov and S. {Di Falco} and E. Diociaiuti and S. Donati and R. Donghia and B. Echenard and C. Ferrari and S. Giovannella and V. Glagolev and F. Grancagnolo and D. Hampai and F. Happacher and D. Hitlin and M. Martini and S. Miscetti and T. Miyashita and L. Morescalchi and P. Murat and D. Pasciuto and E. Pedreschi and G. Pezzullo and F. Porter and F. Raffaelli and A. Saputi and I. Sarra and F. Spinella and G. Tassielli and V. Tereshchenko and Z. Usubov and I.I. Vasilyev and A. Zanetti and R.Y. Zhu}
}

@article{bes3,
doi = {10.1088/1742-6596/293/1/012002},
url = {https://dx.doi.org/10.1088/1742-6596/293/1/012002},
year = {2011},
month = {apr},
publisher = {},
volume = {293},
number = {1},
pages = {012002},
author = {Weiguo Li and (representing BESIII Collaboration)},
title = {{EM} {Calorimeter} in {BESIII} {Experiment}},
journal = {J. Phys.: Conf. Ser.}
}

@article{crystal_ball,
year = {1980},
publisher = {SLAC-R-236},
pages = {Appendix D},
author = {M. J. Oreglia},
title = "{A Study of the Reactions $\psi^{\prime}\to\gamma\gamma\psi$}",
journal = {Ph.D. Thesis}
}

@Inbook{Fabjan2020,
author="Fabjan, C. W. and Fournier, D.",
editor="Fabjan, Christian W. and Schopper, Herwig",
title="Calorimetry",
bookTitle="Particle Physics Reference Library: Volume 2: Detectors for Particles and Radiation",
year="2020",
publisher="Springer International Publishing",
address="Cham",
pages="201--280",
isbn="978-3-030-35318-6",
doi="10.1007/978-3-030-35318-6_6",
url="https://doi.org/10.1007/978-3-030-35318-6_6"
}

@article{pdg,
    author = {Workman, R L and others},
    title = "{Review of Particle Physics}",
    journal = {Prog. Theor. Exp. Phys.},
    volume = {2022},
    number = {8},
    pages = {083C01},
    year = {2022},
    month = {08},
    issn = {2050-3911},
    doi = {10.1093/ptep/ptac097},
    url = {https://doi.org/10.1093/ptep/ptac097},
}

@article{geant4,
title = "{Geant4—a simulation toolkit}",
journal = {Nucl. Instr. Meth. Phys. Res. A},
volume = {506},
number = {3},
pages = {250-303},
year = {2003},
issn = {0168-9002},
doi = {10.1016/S0168-9002(03)01368-8},
url = {https://www.sciencedirect.com/science/article/pii/S0168900203013688},
author = {S. Agostinelli and J. Allison and K. Amako and J. Apostolakis and H. Araujo and P. Arce and M. Asai and D. Axen and S. Banerjee and G. Barrand and F. Behner and L. Bellagamba and J. Boudreau and L. Broglia and A. Brunengo and H. Burkhardt and S. Chauvie and J. Chuma and R. Chytracek and G. Cooperman and G. Cosmo and P. Degtyarenko and A. Dell'Acqua and G. Depaola and D. Dietrich and R. Enami and A. Feliciello and C. Ferguson and H. Fesefeldt and G. Folger and F. Foppiano and A. Forti and S. Garelli and S. Giani and R. Giannitrapani and D. Gibin and J.J. {Gómez Cadenas} and I. González and G. {Gracia Abril} and G. Greeniaus and W. Greiner and V. Grichine and A. Grossheim and S. Guatelli and P. Gumplinger and R. Hamatsu and K. Hashimoto and H. Hasui and A. Heikkinen and A. Howard and V. Ivanchenko and A. Johnson and F.W. Jones and J. Kallenbach and N. Kanaya and M. Kawabata and Y. Kawabata and M. Kawaguti and S. Kelner and P. Kent and A. Kimura and T. Kodama and R. Kokoulin and M. Kossov and H. Kurashige and E. Lamanna and T. Lampén and V. Lara and V. Lefebure and F. Lei and M. Liendl and W. Lockman and F. Longo and S. Magni and M. Maire and E. Medernach and K. Minamimoto and P. {Mora de Freitas} and Y. Morita and K. Murakami and M. Nagamatu and R. Nartallo and P. Nieminen and T. Nishimura and K. Ohtsubo and M. Okamura and S. O'Neale and Y. Oohata and K. Paech and J. Perl and A. Pfeiffer and M.G. Pia and F. Ranjard and A. Rybin and S. Sadilov and E. {Di Salvo} and G. Santin and T. Sasaki and N. Savvas and Y. Sawada and S. Scherer and S. Sei and V. Sirotenko and D. Smith and N. Starkov and H. Stoecker and J. Sulkimo and M. Takahata and S. Tanaka and E. Tcherniaev and E. {Safai Tehrani} and M. Tropeano and P. Truscott and H. Uno and L. Urban and P. Urban and M. Verderi and A. Walkden and W. Wander and H. Weber and J.P. Wellisch and T. Wenaus and D.C. Williams and D. Wright and T. Yamada and H. Yoshida and D. Zschiesche}
}

@techreport{bib:apd,
title = "{8664-1010 Si APD Datasheet}",
author = {Hamamatsu}
}

@article{preshower,
title = {Performance of preshower and shower-maximum detectors with a lead/plastic-scintillator calorimeter},
journal = {Nucl. Instr. Meth. Phys. Res. A},
volume = {487},
number = {3},
pages = {275-290},
year = {2002},
issn = {0168-9002},
doi = {10.1016/S0168-9002(01)00890-7},
url = {https://www.sciencedirect.com/science/article/pii/S0168900201008907},
author = {K Kawagoe and Y Sugimoto and A Takeuchi and T Asakawa and J.P Done and Y Fujii and K Furukawa and F Kajino and T Kamon and N Kanaya and J Kanzaki and S Kim and A Nakagawa and M Nozaki and R Oishi and T Ota and H Takeda and T Takeshita and S Uozumi}
}

@article{segmented_shashlik,
    author = "Li, Linmao and Li, Yulei and Ran, Xinchi and Deng, Zhi and Han, Dong and Wang, Yi",
    title = "{The study of a new longitudinal segmented shashlik electromagnetic calorimeter}",
    doi = "10.1088/1748-0221/20/06/P06033",
    journal = "JINST",
    volume = "20",
    number = "06",
    pages = "P06033",
    year = "2025"
}

\end{document}